%% file: main.tex
\documentclass[a4paper]{spie}  %>>> use this instead for A4 paper
\usepackage{amsmath,amsfonts,amssymb}

\usepackage{graphicx}
\usepackage[colorlinks=true, allcolors=blue]{hyperref}
\usepackage{textgreek} % For using \textmu
\usepackage{support-caption}
\usepackage{subcaption}
\usepackage{rotating}
\usepackage{multirow}
\usepackage{aasmacros}
\usepackage{booktabs}
\usepackage{siunitx}
\usepackage{placeins}
\usepackage[normalem]{ulem}
\usepackage[printonlyused]{acronym}
\usepackage{epigraph}
\usepackage{tabularray}
\usepackage{wasysym}

\newcommand\degree{\hbox{$^\circ$}}

\title{Vector Apodizing Phase Plates for the ELT: From prototype to final optics for METIS and MICADO}
\author[1]{J.A.~van~den~Born}
\author[2]{R.~Landman}
\author[2,3]{D.S.~Doelman}
\author[2]{F.~Snik}
\author[4]{Y.~Nishie}
\author[4]{Y.~Watanabe}
\author[4]{M.~Shoda}
\author[1,2]{F.C.M.~Bettonvil}
\author[1]{E.~Aranzana}
\author[3]{J.H.H.~Rietjens}
\author[2]{T.P.G.~Wijnen}
\author[5]{P.~Baudoz}
\author[6]{O.~Absil}
\author[6]{G.~Orban~de~Xivry}
\author[1]{D.~Dolkens}

\affil[1]{\footnotesize NOVA Optical Infrared Instrumentation Group at ASTRON, Oude Hoogeveensedijk 4, NL-7991 PD Dwingeloo, The Netherlands}
\affil[2]{NOVA, Leiden Observatory, Einsteinweg 55, 2333 CC Leiden, The Netherlands}
\affil[3]{SRON Netherlands Institute for Space Research, Niels Bohrweg 4, 2333 CA Leiden, The Netherlands}
\affil[4]{ColorLink Japan, Ltd., Niigata, Japan}
\affil[5]{LIRA, Observatoire de Paris, Université PSL, CNRS, Sorbonne Université, Université Paris Cité, 5 place Jules Janssen, 92195 Meudon, France}
\affil[6]{STAR Institute, Université de Liège, Allée du Six Août 19c, B-4000 Liège, Belgium}
\authorinfo{E-mail: born@astron.nl}

\begin{document} 
\maketitle

\begin{abstract}
The first generation of instruments for the upcoming Extremely Large Telescope (ELT) will allow for the direct imaging of exoplanets that were previously below the sensitivity or resolution limits of existing facilities through various different High Contrast Imaging capabilities. Both METIS and MICADO will feature one or more Vector Apodizing Phase Plates (vAPP), a type of pupil-plane coronagraph based on liquid crystal technology that allows for broadband phase modification to create zones of high contrast around an observed point source. The METIS vAPPs will operate at \textit{L}- and \textit{M}-band wavelengths between 3.1 and 5.1~\textmu m, while the MICADO vAPP is optimized for shorter wavelengths in the \textit{J}, \textit{H} and \textit{K\textsubscript{s}} bands between 1.15 and 2.32~\textmu m.

In this work, we will provide a brief introduction to this type of coronagraph, followed by a discussion of the commonalities and differences between the METIS and MICADO vAPP designs. In the last year, the final optics have been in production. During preparations for the manufacturing and also during the production phase, various challenges were encountered related to coating inclusions, uniformity of the optically active layers and adhesion between the substrates. Through simulations and empirical findings, we argue that the effects of these imperfections on the on the final optical performance are limited. We will present the expected contrast curves, discuss the current status and reflect on the implications for on-sky observations with METIS and MICADO.
\end{abstract}


\input{acronyms.tex}
\input{01_introduction}

\input{02_manufacturing}
\input{03_performance_indicators}

\input{04_defects}
\input{05_future_outlook}
% \input{06_conclusions}

% Acknowledgements section
% \acknowledgments % equivalent to \section*{ACKNOWLEDGMENTS}       
% \notetoself{Anything or anyone I should mention?}
% \notetoself{Anything or anyone that I should mention?}

%\FloatBarrier
% Appendices
%\newpage
%\appendix    %>>>> this command starts appendixes
% \input{07_appendices}

% References
% \newpage
\bibliography{report} % bibliography data in report.bib
\bibliographystyle{spiebib} % makes bibtex use spiebib.bst

\end{document}

%% file: acronyms.tex
%This document employs several abbreviations and acronyms to refer concisely to an item, after it has been introduced. A complete list of all common acronyms and abbreviations used within the METIS project is provided in \cite{METIS_acronyms}. The following list is aimed at helping the reader in recalling the extended meaning of each short expression of those acronyms appearing in this document:

% List of acronyms included, based on PDR documentation package. Most of the abbreviations used in
% the PDR document package are included, as well as from ESO's Common Definition and Acronyms 
% document (ESO-193178v6, 09-07-2013).
% There are several acronyms that have multiple meanings. Uncomment/Comment the one that you need.
% For an acronym inside another acronym, e.g. MICADO = Multi-AO ..., use
% \acro{MICADO}{Multi-\acs{AO} ...} and label the definition of AO using
% \acro{AO}{Adaptive Optics}\label{acro:AO}.
% Otherwise, warnings are raised.
%\begin{acronym}[MICADO123]
    % A
    \newacro{ACS}{ALMA Common Software}
    \newacro{ACU}{Astrometric Calibration Unit}
    \newacro{AD}{Applicable Documents}\label{acro:AD}
    \newacro{ADC}{Atmospheric Dispersion Corrector}\label{acro:ADC}
    %\newacro{ADC}{Analog to Digital Converter}
    \newacro{ADI}{Angular Differential Imaging}\label{acro:ADI}
    \newacro{ADP}{Acceptance Data Package}
    \newacro{ADU}{Analog-to-Digital Unit}\label{acro:ADU}
    \newacro{AFM}{Active Folding Mirror}
    \newacro{AHU}{Air Handling Unit}
    \newacro{A-IMA}{Astrometric Imaging}
    \newacro{AIV}{Assembly, Integration, Verification}
    \newacro{AIT}{Assembly, Integration, Tests}
    \newacro{AKF}{Auto-collimation Telescope}
    \newacro{ALMA}{Atacama Large Millimeter Array}
    \newacro{ALT}{Altitude}
    \newacro{AMSL}{Above Mean Sea Level}
    \newacro{ANF}{Antofagasta}
    \newacro{AO}{Adaptive Optics}\label{acro:AO}
    \newacro{AOB}{Any Other Business}
    \newacro{AOWFC}{Adaptive Optics Wavefront-sensor Camera}
    \newacro{APE}{Active Phasing Experiment}
    \newacro{API}{Application Programming Interface}
    \newacro{APP}{Apodizing Phase Plate}\label{acro:APP}
    \newacro{AR}{Anti Reflection}\label{acro:AR}
    \newacro{ARR}{Acceptance Readiness Review}\label{acro:ARR}
    \newacro{ARU}{Adaptive Relay Unit}
    \newacro{ASM}{Astronomical Site Monitor}
    %\newacro{ASM}{APE Segmented Mirror}
    \newacro{ASL}{Above Sea Level}
    \newacro{AT}{Alignment Telescope}
    %\newacro{AT}{Acceptance Test}
    \newacro{AWG}{Astrometry Working Group}
    \newacro{AZ}{Azimuth}
    
    % B
    % \newacro{BB}{Blackbody}
    \newacro{BB}{Broad-band (filter}
    \newacro{BMS}{Building Management System}
    \newacro{BOM}{Bill of Materials}
    
    % C
    \newacro{CA}{Cryostat Assembly}
    \newacro{CAD}{Computer Aided Design}\label{acro:CAD}
    \newacro{CAE}{Computer Aided Engineering}
    % \newacro{CAM}{Camera optics}
    \newacro{CAM}{Cold Astrometric Mask}
    \newacro{CBS}{Cost Breakdown Structure}
    \newacro{CCB}{Configuration Control Board}
    % \newacro{CCB}{Change Control Board}
    \newacro{CCD}{Charge Coupled Device}\label{acro:CCD}
    \newacro{CCS}{Central Control System}\label{acro:CCS}
    \newacro{CDR}{Critical Design Review}
    \newacro{CDRL}{Contract Data Requirements List}
    \newacro{CDS}{Correlated Double Sampling}
    \newacro{CFD}{Computational Fluid Dynamics}
    \newacro{CFP}{Call for Proposals}
    \newacro{CFT}{Call for Tender}
    \newacro{CI}{Configuration Item}
    \newacro{CIDL}{Configuration Item Data List}
    \newacro{CIL}{Critical Item List}
    \newacro{C-IMA}{Coronagraphic Imaging}
    \newacro{CIR}{Central Intensity Ratio}
    \newacro{CM}{Configuration Management}
    \newacro{CLC}{Clasical Lyot Coronograph}
    \newacro{CLIP}{C Library for Image Processing}
    \newacro{CLJ}{ColorLink Japan, Ltd.}\label{acro:CLJ}
    \newacro{CMM}{Coordinate Measurement Machine}\label{acro:CMM}
    \newacro{CMMS}{Computerized Maintenance Management System}
    \newacro{CMP}{Chemical Mechanical Polishing}
    \newacro{CMS}{Cabinet Management System}
    \newacro{COG}{Center of Gravity}
    \newacro{COI}{Cold Optics Instrument}
    \newacro{COL}{Collimator}
    \newacro{COLR}{Collimator Reference Plate}
    \newacro{Co-Pi}{Co-Principal Investigator}
    \newacro{COS}{Coordinate System}\label{acro:CS}
    \newacro{COTS}{Component Of The Shelf}
    \newacro{CPL}{Common Pipeline Library}
    \newacro{CPU}{Central Processing Unit}
    \newacro{CQC}{Complete Quadratic mode Combination}
    \newacro{CRB}{Change Request Board}
    \newacro{CRDL}{Contract Requirement Data List}
    \newacro{CRE}{Change Request}
    \newacro{CRiSP}{CRyostat in SPectrometer}\label{acro:CRiSP}
    \newacro{CTB}{Cryogenic Test Bench}\label{acro:CTB}
    \newacro{CTE}{Coefficient of Thermal Expansion}\label{acro:CTE}
    \newacro{CVD}{Chemical Vapor Deposition}
    \newacro{CVI}{Chemical Vapor Infiltration}
    \newacro{CWM}{Central Wheel Mechanism}
    
    % D
    \newacro{DB}{Database}
    \newacro{DBCM}{Database Content Management}
    \newacro{DCL}{Declared Component List}
    \newacro{DCS}{Detector Control System}
    \newacro{DD}{Deliverable Document}
    \newacro{DDF}{Design Definition File}
    \newacro{DEC}{Declination}
    \newacro{DFS}{Data Flow Subsystem}
    \newacro{DICB}{Data Interface Control Board}
    \newacro{DID}{Data Item Description}
    \newacro{DIMM}{Differential Image Motion Monitor}
    \newacro{DIN}{Deutsche Industrie Norm}
    \newacro{DIPSI}{Diffraction Image Phase Sensing Instrument}
    \newacro{DIT}{Detector Integration Time}
    \newacro{DJF}{Design Justification File}
    \newacro{DLR}{Damage Limitation Requirement}
    \newacro{DM}{Deformable Mirror}
    \newacro{DMPPL}{Declared Materials, Parts and Process List}
    \newacro{DOF}{Degrees Of Freedom}
    \newacro{DPS}{Data Processing Subsystem}
    % \newacro{DPS}{Detector Positioning System}
    \newacro{DRB}{Data Reduction Block}
    \newacro{DRD}{Document Requirements Definition}
    \newacro{DR}{De-Rotator}
    \newacro{DRLS}{Data Reduction Library Specification document}
    \newacro{DRS}{Data Reduction System}
    
    % E
    \newacro{EC}{European Commission}
    \newacro{ECSS}{European Cooperation for Space Standardization}
    \newacro{EEE}{Electrical Electornic \& Electromechanical components}
    \newacro{EELT}{European Extremely Large Telescope}
    \newacro{E-ELT}{European Extremely Large Telescope}
    \newacro{EH}{Entrance Hall}
    \newacro{ELGS}{\acs{ELT} Laser Guide Star}
    \newacro{ELT}{Extremely Large Telescope}\label{acro:ELT}
    \newacro{ELT DS}{\acs{ELT} Design Study}
    \newacro{EMC}{Electromagnetic Compatibility}
    \newacro{EN}{Euro Norm}
    \newacro{ENC}{Enclosure}
    \newacro{EPICS}{Exo-Planets Imaging Camera and Spectrograph}
    \newacro{ERA}{European Research Area}
    \newacro{ERIS}{Enhanced Resolution Imager and Spectrograph}\label{acro:ERIS}
    \newacro{ES}{Edge Sensor}
    \newacro{ESE}{\acs{ELT} Science \& Engineering (committee)}
    \newacro{ESFRI}{European Strategy Forum on Research Infrastructures}
    \newacro{ESI}{Edge Sensor Interface}
    \newacro{ESO}{European Southern Observatory}
    \newacro{ETC}{Exposure Time Calculator}
    \newacro{ETH}{Eidgenössische Technische Hochschule Zürich}\label{acro:ETH}
    \newacro{ETP}{\acs{ELT} Telescope Project Office}
    \newacro{EUV}{Extreme Ultraviolet}
    \newacro{EXW}{EX Works}
    
    % F
    \newacro{FCA}{Free Carrier}
    \newacro{FCB}{Friction Characterization Breadboard}
    \newacro{FCU}{Flat field and spectral Calibration Unit}
    \newacro{FCS}{Function Control Software}
    \newacro{FDIR}{Failure Detection, Isolation and Recovery}
    \newacro{FDR}{Final Design Review}
    \newacro{FE}{Finite Element}
    \newacro{FEA}{Finite Element Analysis}
    \newacro{FEED}{Front End Engineering Design}
    \newacro{FEM}{Finite Element Model}
    \newacro{FFT}{Fast Fourier Transform}
    \newacro{FITS}{Flexible Image Transport System}
    \newacro{FLI}{Fractional Lunar Illumination}
    \newacro{FMEA}{Failure Mode and Effect Analyses}
    \newacro{FMECA}{Failure Mode Effect and Criticality Analysis}
    \newacro{FOB}{Free On Board}
    \newacro{FOV}{Field Of View}
    \newacro{FP}{Focal Plane}
    \newacro{FPGA}{Field Programmable Gate Array}
    \newacro{FPA}{Focal Plane Array}
    \newacro{FPM}{Focal Plane Mechanism}
    \newacro{FPn}{European Commission's Framework Programme n (n=1, 2, ...).}
    \newacro{FPP}{Focal Plane Mosaic}
    \newacro{FRACAS}{Failure Reporting, Analysis and Corrective Action System}
    \newacro{FTA}{Failure Tree Analysis}
    \newacro{FTE}{Full Time Equivalent (year)}
    \newacro{FTS}{Fourier Transform Spectrograph}
    \newacro{FW}{Firmware}
    \newacro{FWC}{Full Well Capactiy}
    \newacro{FWM}{Filter Wheel Mechanism}
    \newacro{FWHM}{Full Width at Half Maximum}
    
    % G
    % \newacro{GCS}{General Command Set}
    \newacro{GCS}{Global Coordinate System}
    \newacro{GD}{Green Donut}
    \newacro{GDS}{Green Donut Support structure}
    \newacro{GI}{Gravity Invariant}
    \newacro{GLAO}{Ground Layer Adaptive Optics}
    \newacro{GMT}{Giant Magellan Telescope}
    \newacro{GMT}{Greenwich Meridian Time}
    \newacro{GSE}{Global Support Equipment}
    \newacro{GTO}{Guaranteed Time Observations}
    \newacro{GUI}{Graphical User Interface}
    \newacro{GUID}{Globally Unique Identifier}
    \newacro{gvAPP}{Grating Vector Apodizing Phase Plate}
    
    % H
    \newacro{HB}{Hydrostatic Bearings}
    \newacro{HCI}{High Contrast Imaging}
    \newacro{HDRL}{High-level Data Reduction Library}
    \newacro{HDU}{Header-Data Unit}
    \newacro{HMI}{Human Machine Interface}
    \newacro{HRI}{High Resolution Imager}
    \newacro{HRTC}{Hard Real Time Computer}
    \newacro{HTRA}{High Time Resolution Astrophysics}
    \newacro{HVAC}{Heating, Ventilation and Air Conditioning}
    \newacro{HW}{Hardware}
    
    % I
    \newacro{IAA}{Instrument Assembly Area}
    \newacro{IAC}{Instituto de Astrofísica de Canarias}
    \newacro{IAS}{Internal Alignment System}
    \newacro{ICD}{Interface Control Document}
    \newacro{ICS}{Instrument Control System}
    \newacro{ICT}{Information and Communications Technology}
    \newacro{I/F}{Interface}
    \newacro{IF}{Interface}
    \newacro{IFU}{Integral Field Unit}
    \newacro{ILS}{Interlock System}
    % \newacro{ILS}{Integrated Logistics and Support}
    \newacro{IM}{Integrated Model}
    % \newacro{IM}{Integrated Modeling}
    % \newacro{IM}{Interaction Matrix}
    \newacro{IMF}{Intermediate Focus}
    \newacro{IMG}{Imager, usually refers to the METIS LM imager subsystem.}\label{acro:IMG}
    \newacro{INS}{Instrumentation Software}
    \newacro{INT}{Isaac Newton Telescope}
    \newacro{IR}{Infrared}
    \newacro{IRM}{Integration Readiness Meeting}
    \newacro{IS}{Integrating Sphere}
    \newacro{ISDD}{Instrument Software Design Description}
    \newacro{ISFS}{Instrument Software Functional Specification}
    \newacro{ISO}{International Standards Organisation}
    \newacro{Iss.}{Issue (number)}
    \newacro{ISURS}{Instrument Software User Requirement Specification}
    \newacro{IT}{Information Technology}
    \newacro{IVV}{In Vino Veritas}
    \newacro{IWA}{inner working angle}
    
    % J
    \newacro{JWST}{James Webb Space Telescope}
    
    % K
    \newacro{KIP}{Key Inspection Point}
    \newacro{KM}{Key Milestone}
    \newacro{KOM}{Kick-Off Meeting}
    
    % L
    \newacro{LADA}{\acs{LGS} Adapter nasmyth A focus}
    \newacro{LADB}{\acs{LGS} Adapter nasmyth B focus}
    \newacro{LADC}{Longitudinal Atmospheric Dispersion Corrector}
    \newacro{LAN}{Local Area Network}
    \newacro{LBT}{Large Binocular Telescope}\label{acro:LBT}
    \newacro{LCS}{Local Coordinate System}
    % \newacro{LCS}{Local Control System}
    \newacro{LCP}{Liquid Crystal Polymer}\label{acro:LCP}
    \newacro{LCU}{Local Control Unit}
    \newacro{LESIA}{Laboratoire d'Etudes Spatiales et Instrumentations pour l'Astrophysique}
    \newacro{LEMP}{Lightning Electromagnetic Pulse}
    \newacro{LGS}{Laser Guide Star}\label{acro:LGS}
    \newacro{LGSF}{Laser Guide Star Facility}
    \newacro{LGSU}{Laser Guide Star(s) Unit}
    \newacro{LMS}{LM-band Spectrograph}\label{acro:LMS}
    \newacro{LOR}{Low Order Reference \acs{WFS}}
    \newacro{LoS}{Line of Sight}
    \newacro{LPP}{Linearly Photopolymerizable Polymer}\label{acro:LPP}
    \newacro{LPS}{Lightning Protection System}
    % \newacro{LPS}{Laser Projection Subunit}
    \newacro{LRU}{Line Replaceable Unit}
    \newacro{LSU}{Local Safety Unit}
    \newacro{LSV}{Local Supervisor}
    \newacro{LTAO}{Laser Tomography Adaptive Optics}
    \newacro{LRI}{Low Resolution Imager}
    \newacro{LRU}{Line Replacement Unit}
    \newacro{LUT}{Look Up Table}
    
    % M
    \newacro{M1}{Primary mirror}
    \newacro{M2}{Secondary mirror}
    \newacro{M3}{Tertiary mirror}
    \newacro{M4}{Quaternary mirror}
    \newacro{M5}{Fifth mirror}
    \newacro{M6}{Sixth mirror}
    \newacro{M6C}{Sixth mirror, folding the beams towards the coud\'e focus}
    \newacro{M6G}{Sixth mirror, folding the beams towards the gravity-invariant focus}
    \newacro{M6N}{Sixth mirror, folding the beams towards the lateral (bent) Nasmyth foci}
    \newacro{MandM}[M\&M]{\acs{MAORY} and \acs{MICADO} configuration}
    \newacro{MAIT}{Manufacture, Assembly, Integration, Test}
    \newacro{MAN}{Manual}
    \newacro{MAORY}{Multi-conjugate Adaptive Optics RelaY for the \acs{ELT}}\label{acro:MAORY}
    \newacro{mas}{milli arcsecond}
    \newacro{MASS}{Multi-Aperture Scintillation Sensor}
    \newacro{MBS}{Main Bench Structure}
    \newacro{MCA}{\acs{MICADO} Calibration Assembly}\label{acro:MCA}
    \newacro{MCAO}{Multi Conjugate Adaptive Optics}
    \newacro{MCD}{\acs{MICADO}}
    \newacro{MCU}{Movable source Calibration Unit}
    % \newacro{MCU}{Micro-Controller Unit}
    \newacro{MDU}{\acs{MCA} Deployment Unit}
    \newacro{METIS}{Mid-infrared ELT Imager and Spectrograph}\label{acro:METIS}
    \newacro{MHE}{Mechanical Handling Equipment}
    \newacro{MICADO}{Multi-\acs{AO} Imaging CamerA for Deep Observations}\label{acro:MICADO}
    \newacro{MIMO}{Mulitple Input Multiple Output}
    \newacro{MIP}{Mandatory Inspection Point}
    \newacro{MLE}{Maximum Likely Earthquake}
    \newacro{MMA}{\acs{MICADO} \& \acs{MAORY} Astrometry}
    \newacro{MMB}{Mechanical Maintenance Building}
    \newacro{MMPP}{Materials, Mechanical Parts and Processes}
    \newacro{MOAO}{Multi Object Adaptive Optics}
    \newacro{MPE}{Max-Planck-Institut für extraterrestrische Physik}
    \newacro{MPIA}{Max-Planck-Institut für Astronomie}\label{acro:MPIA}
    \newacro{MRB}{Material Review Board}
    \newacro{MRF}{Magneto Rheological Finishing}
    \newacro{MS}{Maintenance software}
    \newacro{MSE}{Mechanical Support Equipment}
    \newacro{MSM}{Main Selection Mechanism}
    \newacro{MTBCF}{Mean Time Between Critical Failures}
    \newacro{MTBF}{Mean Time Between Failure}
    \newacro{MTF}{Modulation Transfer Function}
    \newacro{MTR}{Multi Twist Retarder}\label{acro:MTR}
    \newacro{MTTF}{Mean Time To Failure}
    \newacro{MTTR}{Mean Time To Repair}
    
    % N
    \newacro{N/A}{Not Applicable}
    \newacro{NA}{Numerical Aperture}
    \newacro{NB}{Narrow-band (filter)}
    \newacro{NADA}{\acs{NGS} Adapter nasmyth A focus}
    \newacro{NADB}{\acs{NGS} Adapter nasmyth B focus}
    \newacro{NADC}{\acs{NGS} Adapter nasmyth C focus}
    % \newacro{NADC}{\acs{NGS} Adapter Coud\'e focus}
    \newacro{NADG}{\acs{NGS} Adapter nasmyth G focus}
    \newacro{NADA}{\acs{NGS} Adapter invariant Gravity focus}
    \newacro{NC}{Non-Compliant, Non-Compliance}
    \newacro{NCPA}{Non Common Path Aberration}
    \newacro{NCR}{Non-Conformance Report}
    % \newacro{NCR}{No Collapse Requirement}
    \newacro{ND}{Neutral Density}
    \newacro{NDA}{Non-Disclosure Agreement}
    \newacro{NGC}{New Generation Detector Controller}
    \newacro{NGS}{Natural Guide Star}\label{acro:NGS}
    \newacro{NIR}{Near Infrared}
    \newacro{NOA61}{Norland Optical Adhesive 61}\label{acro:NOA61}
    \newacro{NOVA}{Nederlandse Onderzoekschool voor Astronomie}\label{acro:NOVA}
    
    % O
    \newacro{OAPD}{Osservatorio Astronomico di Padova}
    \newacro{OB}{Observation Block}
    \newacro{OBE}{Operating Basis Earthquake}
    \newacro{OBS}{Organisational Breakdown Structure}
    \newacro{OCD}{Operational Concept Description}
    \newacro{OCS}{Observation Coordination Software}
    \newacro{OD}{Optical Density}\label{acro:OD}
    \newacro{OLDB}{Online Database}
    \newacro{OoM}{Order of Magnitude}
    \newacro{OpcUA}{OPC Unified Architecture}
    \newacro{OPD}{Optical Path Difference}
    \newacro{OPL}{Optical Path Length}
    \newacro{ORM}{Observatorio Roque de los Muchachos (Canary Islands, Spain)}
    \newacro{OSE}{Optical Support Equipment}
    \newacro{OSS}{Observer Support Software}
    \newacro{OWA}{outer working angle}
    \newacro{OWL}{Overwhelmingly Large telescope}
    
    % P
    \newacro{P2PP}{Phase 2 Proposal Preparation}
    \newacro{PA}{Product Assurance}
    \newacro{PAC}{Preliminary Acceptance in Chile}
    \newacro{PACT}{Position Actuator}
    \newacro{PAE}{Preliminary Acceptance in Europe}
    \newacro{PAL}{Part List}
    \newacro{PAO}{Preliminary Acceptance at the Observatory}
    \newacro{PA/QE}{Product Assurance / Quality Assurance}
    \newacro{PAP}{Product Assurance Plan}
    \newacro{PCA}{Principal Component Analysis}\label{acro:PCA}
    \newacro{PDM}{Product Data Managment}
    \newacro{PDR}{Preliminary Design Review}
    \newacro{PFS}{Pre-Focal Station}
    \newacro{PGA}{Peak ground Acceleration}
    \newacro{PHA}{Preliminary Hazard Analysis}
    \newacro{PI}{Principal Investigator}
    \newacro{PIM}{Pupil Imager}
    \newacro{PLC}{Programmable Logic Controller}
    \newacro{PMP}{Project Management Plan}
    \newacro{PO}{Project Office}
    \newacro{POA}{Paranal Or Armazones}
    \newacro{ppFPA}{Partially Populated Focal Plane Array}
    \newacro{PreCADO}{Observations Preparation Software System}
    \newacro{PRR}{Preliminary Requirements Review}
    \newacro{PS}{Plate Scale}
    \newacro{PSD}{Power Spectral Density}
    \newacro{PSF}{Point Spread Function}\label{acro:PSF}
    \newacro{PSF-R}{\acs{PSF} Reconstruction}
    \newacro{PSI}{Pound per Square Inch}
    % \newacro{PSI}{Point Source Interferometer}
    \newacro{PSS}{Point Source Sensitivity}
    \newacro{PSSn}{Point Source Sensitivity, normalized}
    \newacro{PTP}{Precision Time Protocol}
    \newacro{PTV}{Peak To Valley}
    \newacro{PV}{Peak to Valley}
    \newacro{PVA}{Provisional Acceptance}
    \newacro{PVD}{Phase Vapor Deposition}
    \newacro{PWFS}{Pyramid Wavefront Sensor}
    \newacro{PWM}{Pupil Wheel Mechanism}
    \newacro{PYPS}{Pyramid Phasing Sensor}

    % Q
    \newacro{QA}{Quality Assurance}
    \newacro{QC}{Quality Control}
    \newacro{QE}{Quantum Efficiency}
    \newacro{QoS}{Quality of Service}
    \newacro{Qty}{Quantity}
    
    % R
    \newacro{RA}{Right Ascension}
    \newacro{RAM}{Reliability, Availability, Maintainability}
    \newacro{RAMS}{Reliability, Availability, Maintainability, Safety}
    \newacro{RCA}{Root Cause Analysis}
    \newacro{RandD}[R\&D]{Research \& Development}
    \newacro{RD}{Reference Documents}
    \newacro{RFP}{Request For Proposal}
    \newacro{RFW}{Request For Waiver}
    \newacro{RIC}{Review Item Comment}
    \newacro{RID}{Review Item Discrepancy}
    \newacro{RIQ}{Review Item Question}
    \newacro{RM}{Remote Mode}
    \newacro{RMS}{Root Mean Square}
    \newacro{RO}{Relay Optic}
    \newacro{RON}{Read Out Noise}
    \newacro{RPC}{Remote Procedure Call}
    \newacro{RRM}{Rapid Response Module}
    \newacro{RRR}{Read-Reset-Read}
    \newacro{RSS}{Root Sum Square}
    \newacro{RTC}{Real Time Computer}
    \newacro{RTD}{Real Time Display}
    
    % S
    \newacro{SA}{Support Astronomer}
    \newacro{SACOS}{Segment Active Optics Sensor}
    \newacro{SAF}{Science Archive Facility}
    \newacro{SAM}{Sparse Aperture Mask}
    \newacro{SCAO}{Single Conjugate Adaptive Optics}\label{acro:SCAO}
    \newacro{SCIDAR}{Scintillation Detection And Ranging}\label{acro:SCIDAR}
    \newacro{SCL}{Santiago de Chile}
    \newacro{SCP}{Service Connection Point}
    \newacro{SCU}{\acs{SCAO} Calibration Unit}
    \newacro{SE}{System Engineering}
    \newacro{SED}{Spectral Energy Distribution}
    \newacro{SFP}{Scientific Focal Plane}
    \newacro{SFS}{Segment Figure Sensor}
    \newacro{SHAPS}{Shack-Hartmann Phasing Sensor}
    \newacro{SiC}{Silicon Carbide}
    \newacro{S-IMA}{Standard Imaging}
    \newacro{SimCADO}{Instrument Data Simulator for \acs{MICADO}}
    \newacro{SISO}{Single Input Single Output}
    \newacro{SLODAR}{Slope Detection And Ranging}
    \newacro{SM}{Service Mode}
    \newacro{SNR}{Signal to Noise Ratio}
    \newacro{SOO}{Statement Of Objectives}
    \newacro{SOW}{Statement Of Work}
    \newacro{SPD}{Surge Protective Mode}
    \newacro{SPE}{Spectroscopy}
    % \newacro{SPE}{Specification}
    \newacro{SPF}{Single Point Failure}
    \newacro{SR}{Strehl Ratio}
    \newacro{SRD}{System Requirements Document}
    \newacro{SRON}{Space Research Organisation Netherlands}\label{acro:SRON}
    \newacro{SRTC}{Soft Real Time Computer}
    \newacro{SRU}{Shop Replaceable Unit}
    \newacro{SSO}{Solar System Object}
    \newacro{SSS}{Single Star \acs{SCIDAR}}
    \newacro{ST}{Support Structure}
    \newacro{SUP}{Supervisor}
    \newacro{SV}{Science Verification}
    \newacro{SW}{Software}
    \newacro{SWPA}{Software Product Assurance}
    
    % T
    \newacro{T-AWG}{Taskforce Astrometry Working Group}
    \newacro{TACOS}{Telescope Active Optics Sensor}
    \newacro{TBC}{To Be Confirmed}\label{acro:TBC}
    \newacro{TBD}{To Be Determined}\label{acro:TBD}
    \newacro{TBW}{To Be Written}
    \newacro{TCS}{Telescope Control System}
    \newacro{TDCS}{(Technical) Detector Control Software}
    \newacro{TDP}{Technical Data Package}
    \newacro{THD}{Total Harmonic Distortion}
    \newacro{TIO}{Telescope and Instrument Operator}
    \newacro{TLI}{Threshold Limited Integration}
    \newacro{TLR}{Top Level Requirements}
    \newacro{TMA}{Three Mirror Anastigmat}
    \newacro{TMT}{Thirty Meter Telescope}
    \newacro{TN}{Technical Note}
    \newacro{TOO}{Target Of Opportunity}
    \newacro{TRE}{Technology Readiness Level}
    \newacro{TRL}{Technology Readiness Level}
    \newacro{TRM}{Test Readiness Meeting}
    \newacro{TRR}{Test Readiness Review}\label{acro:TRR}
    \newacro{TRS}{Time Reference System}
    \newacro{TS}{Technical Specification}
    \newacro{TSS}{Top Support Structure}
    \newacro{TT}{Tip-Tilt}
    
    % U
    \newacro{UKA}{\acs{UKATC}}
    \newacro{UKATC}{United Kingdom Astronomical Technology Centre}\label{acro:UKATC}
    \newacro{UPS}{Uninterruptible Power Supply}
    \newacro{URD}{User Requirements Document}
    \newacro{USM}{Universitäts-Sternwarte München}
    \newacro{USP}{User defined Polarizing Surface}
    % \newacro{UT}{Universal Time}
    \newacro{UT}{Unit Telescope}
    \newacro{UTC}{Coordinated Universal Time}
    \newacro{UV}{Ultraviolet}
    
    % V
    \newacro{vAPP}{Vector Apodizing Phase Plate}
    \newacro{VCD}{Verification Control Document}
    \newacro{ver.}{version}
    \newacro{VIS}{Visible light}
    \newacro{VLT}{Very Large Telescope}
    \newacro{VLTI}{Very Large Telescope Interferometer}
    \newacro{VM}{Virtual Machine}
    % \newacro{VM}{Visitor Mode}
    \newacro{VO}{Virtual Observatory}
    \newacro{VODCA}{Vortex Optical Demonstrator for Coronagraphic Application}\label{acro:VODCA}
    \newacro{VPM}{Vortex Phase Mask}\label{acro:VPM}
    \newacro{vs.}{versus}
    \newacro{VV}[V\&V]{Validation and Verification}
    \newacro{VVC}{Vector Vortex Coronagraph}
    
    % W
    \newacro{WAM}{Warm Astrometric Mask}
    \newacro{WBS}{Work Breakdown Structure}
    \newacro{WEB}{Wind Evaluation Experiment}
    \newacro{WF}{Wavefront}
    \newacro{WFC}{Wavefront Control}
    \newacro{WFE}{Wavefront Error}
    \newacro{WFS}{Wave Front Sensor}\label{acro:WFS}
    \newacro{WFRTC}{Wavefront Real Time Computer}\label{acro:WFRTC}
    \newacro{WH}{Warping Harness}
    \newacro{WHT}{William Herschel Telescope}
    \newacro{WP}{Work Package}
    \newacro{WPD}{Work Package Definition}
    \newacro{w.r.t.}{with respect to}
    
    % X
    \newacro{X4}{MPE integration hall}
    \newacro{XAO}{Extreme Adaptive Optics}
    
    % Y
    
    % Z
    \newacro{ZEUS}{Zernike Unit (segment phasing) Sensor}
    
%\end{acronym}

%% file: 01_introduction.tex
\section{Introduction}
The study of protoplanetary disks and planet formation and the detection and characterization of exoplanets through direct imaging are among the primary science objectives of \acs{METIS}\cite{Brandl2022, Brandl2026} and \acs{MICADO}\cite{Davies2018, Davies2026}, made possible by the large collecting area, high sensitivity and exquisite spatial resolution that will be offered by the 39 meter primary mirror of the \acs{ELT}. Generally, to obtain direct images from one or multiple faint sources close to a bright star, the light from the observed star is suppressed or locally removed from the image to reveal a weak companion point source (an exoplanet) or an extended region with low surface brightness (a circumstellar disk)\cite{Doelman2022,Follette2023,Kenworth_Haffert_HCI_review_2025,Maio2025}. Both first generation \acs{ELT} instruments will implement multiple \ac{HCI} capabilities\cite{Absil_METIS_HCI, Baudoz_MICADO_HCI}, including classical Lyot coronagraphs, vortex coronagraphs, sparse aperture masking and vector Apodizing Phase Plates (\acs{vAPP}), to obtain raw contrast levels around 10\textsuperscript{-5}. This proceeding will focus on the latter, the \acs{vAPP}, a pupil optic that uses polarization sensitive phase modulation to create a local region with very low amounts of starlight\cite{Snik2012}.

\begin{figure}
    \centering
    \begin{minipage}[t]{.49\textwidth}
        \captionsetup{width=0.95\linewidth}
        \includegraphics[width=\linewidth]{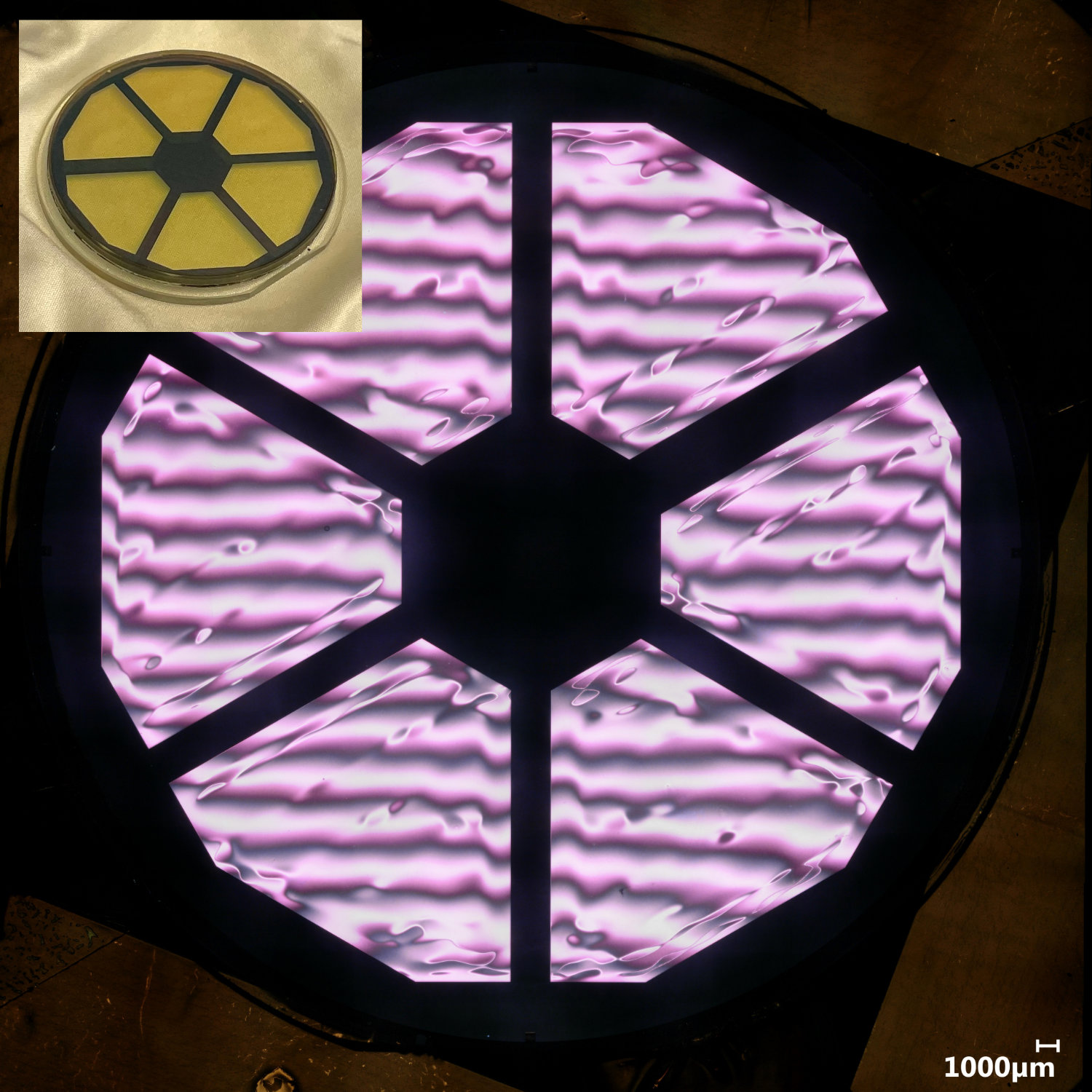}
        \captionof{figure}{The manufactured METIS \acs{vAPP} for the METIS imager as seen through cross-polarizers. The inset shows the optic as it looks under normal room lighting. One of the six arms in the pattern is thicker to enable focal plane wavefront sensing\cite{Orban2024,Bos2019,Bos2021}. The patterned area is approximately 44~mm in diameter.}
        \label{fig:APP-IMG}
    \end{minipage}
    \begin{minipage}[t]{.49\textwidth}
        \captionsetup{width=0.95\linewidth}
        \includegraphics[width=\linewidth]{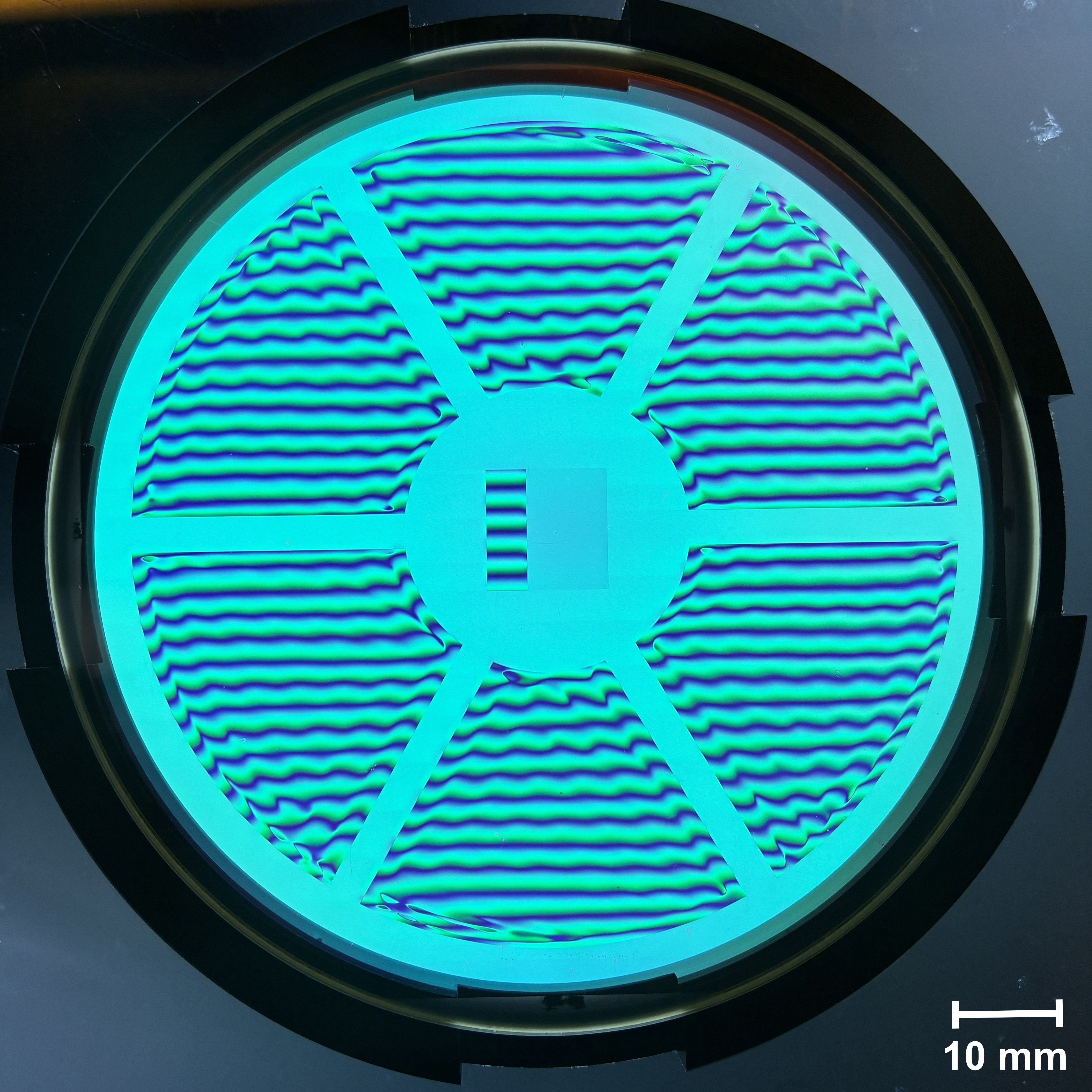}
        \captionof{figure}{The \acs{vAPP} for MICADO at the end of production, before lamination with the second substrate. Here too, shown through cross polarizing sheets. The pattern in the middle was used for monitoring the coating progression and will be covered by the amplitude mask later. The patterned area is approximately 75~mm in diameter.}
        \label{fig:APP-MICADO}
    \end{minipage}
\end{figure}

Recently, two of the three \acs{vAPP}s planned for the first generation of \ac{ELT} instruments have been manufactured by \acl{CLJ}, in close collaboration with \acs{NOVA}. As of June 2026, the \acs{vAPP} for the METIS imager\cite{METIS_IMG2024}, from here on referred to as the APP-IMG, has been delivered to the instrument team. The second \acs{vAPP} for METIS, which will be integrated into the \textit{L} and \textit{M}-band spectrograph of the instrument (hereafter, APP-LMS), will start production in the coming months. Finally, the MICADO \acs{vAPP} has been manufactured and now awaits final lamination. The two manufactured optics are shown in Figs.~\ref{fig:APP-IMG} and \ref{fig:APP-MICADO}. First, we will describe the operating principle and design of these type of coronagraphs. In the following sections, we will summarize the manufacturing process and provide results of the key performance indicators after prototyping the required liquid crystal recipes. Then, in section~\ref{sec:defects}, we will present in detail the manufacturing challenges we have encountered, including coating defects and large wavefront errors. We end with a brief outlook, together with our conclusions.

\subsection{Operating principle of the (gv)APP}
A detailed overview and in-depth review of the different types of \ac{APP}s is given in Ref.~\citenum{Doelman2021}. In summary, an \acs{APP} works by modifying the phase in the optical pupil to diffract light away from a region around the imaged point source to create an area of exceptionally low intensity. This so called dark zone is well suited for the detection of a faint companion to an observed host star. Although the planet image is also affected by the pupil optic, the increase in contrast offered by the creation of a dark zone is larger than the decrease in planet signal.

Initially, pupil phase apodization was done with diamond turned phase plates\cite{Kenworthy2007}. Because these are inherently chromatic, a major improvement was obtained by switching to self-aligning and birefringent \ac{LCP} and exploiting the concept of geometric phase for circularly polarized light\cite{Escuti2016}. Here, multiple layers are stacked to create a broadband half-wave retarder and the phase pattern is manipulated through a direct-writing (or maskless photolithography) process that sets the local fast axis orientation. For a half-wave plate, the phase angle $\phi$ is equal to twice the fast axis orientation $\theta$, and it changes in sign corresponding to the circular polarization state.  \acs{APP}s using these birefringent liquid crystals are called \acl{vAPP}s (\acs{vAPP})\cite{Snik2012}. An observation of an unpolarized point source done with a \acs{vAPP} results in two superimposed PSF images with opposite circular polarization. A polarization grating is added to the fast axis pattern to achieve spatial separation of the two PSF images and thus prevents light from one polarization state leaking into the dark zone of the other. A \acs{vAPP} that includes a grating pattern is sometimes referred to as a \ac{gvAPP}\cite{Otten2014}.

Because of fundamental material properties and manufacturing inaccuracies, the efficiency of any produced half-wave plate is never perfect. For the \acs{vAPP}s this has the effect that some light will leak from the first order diffraction and appear at the zeroth order diffraction location. This undesired leakage term can prove useful in practice as a photometric and astrometric reference during \acs{HCI} observations, because the two circularly polarized copies of the \acs{PSF} can easily be saturated when observing a bright target, even at maximum detector read-out speeds. Typically, the leakage term is of the order of a few per cent of the total starlight incident upon the \acs{vAPP}. %Although a bright core remains in both PSF cores -- this intensity peak is critical; it is the signal that is picked up as the faint companion in the observation -- this signal is often saturated and can therefore not be used to extract useful photometric and astrometric information about the observed source. The leakage term is of much lower intensity and therefore ideally suited for this purpose. Typically, the leakage term is of the order of a few per cent of the total starlight incident upon the \acs{vAPP}.

\subsection{The design of the METIS and MICADO \acl{vAPP}s}
\begin{figure*}[tb!]
    \centering
    \begin{subfigure}[c]{0.3\textwidth}
        \centering
        \includegraphics[width=\linewidth]{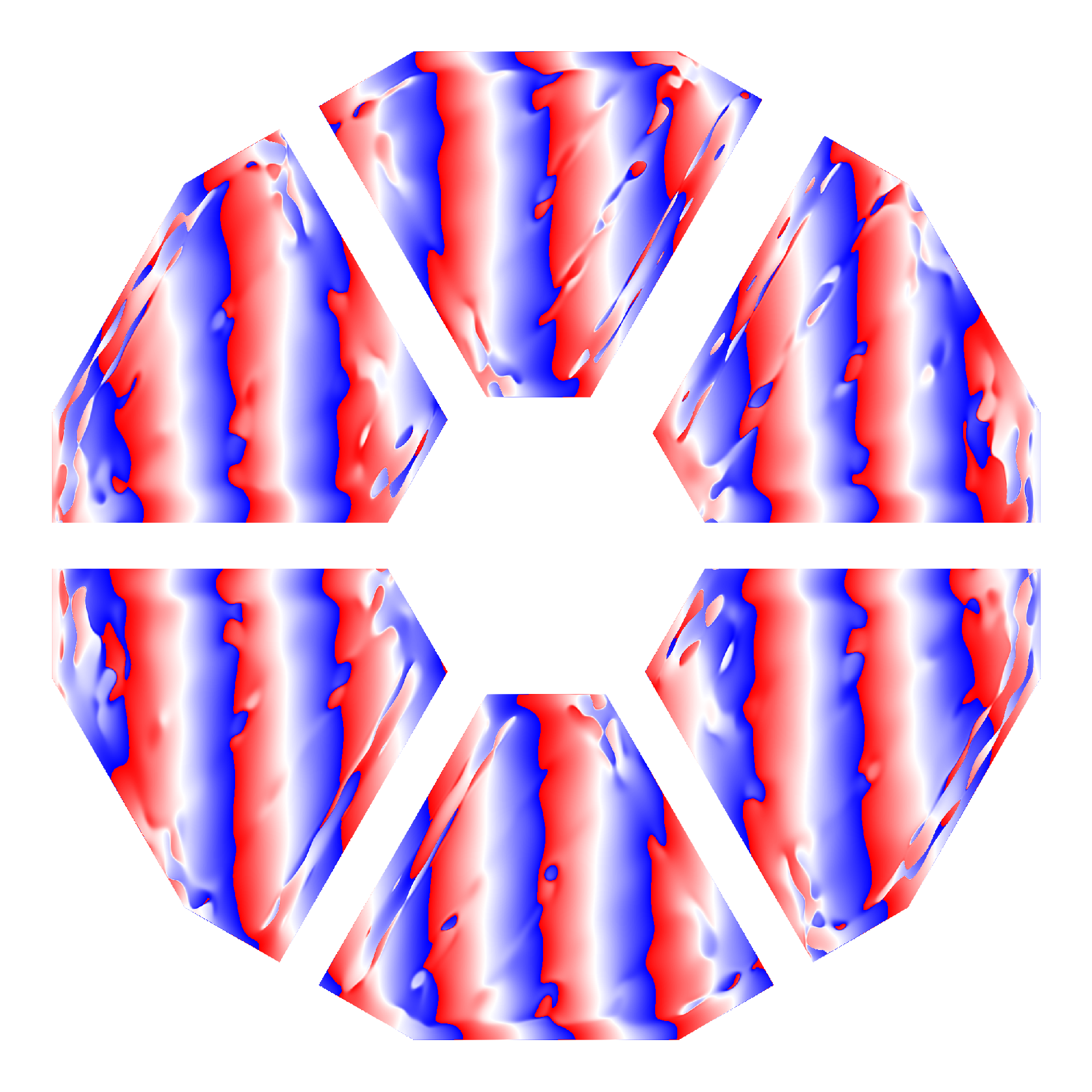}
    \end{subfigure}%
    ~ 
    \begin{subfigure}[c]{0.3\textwidth}
        \centering
        \includegraphics[width=\linewidth]{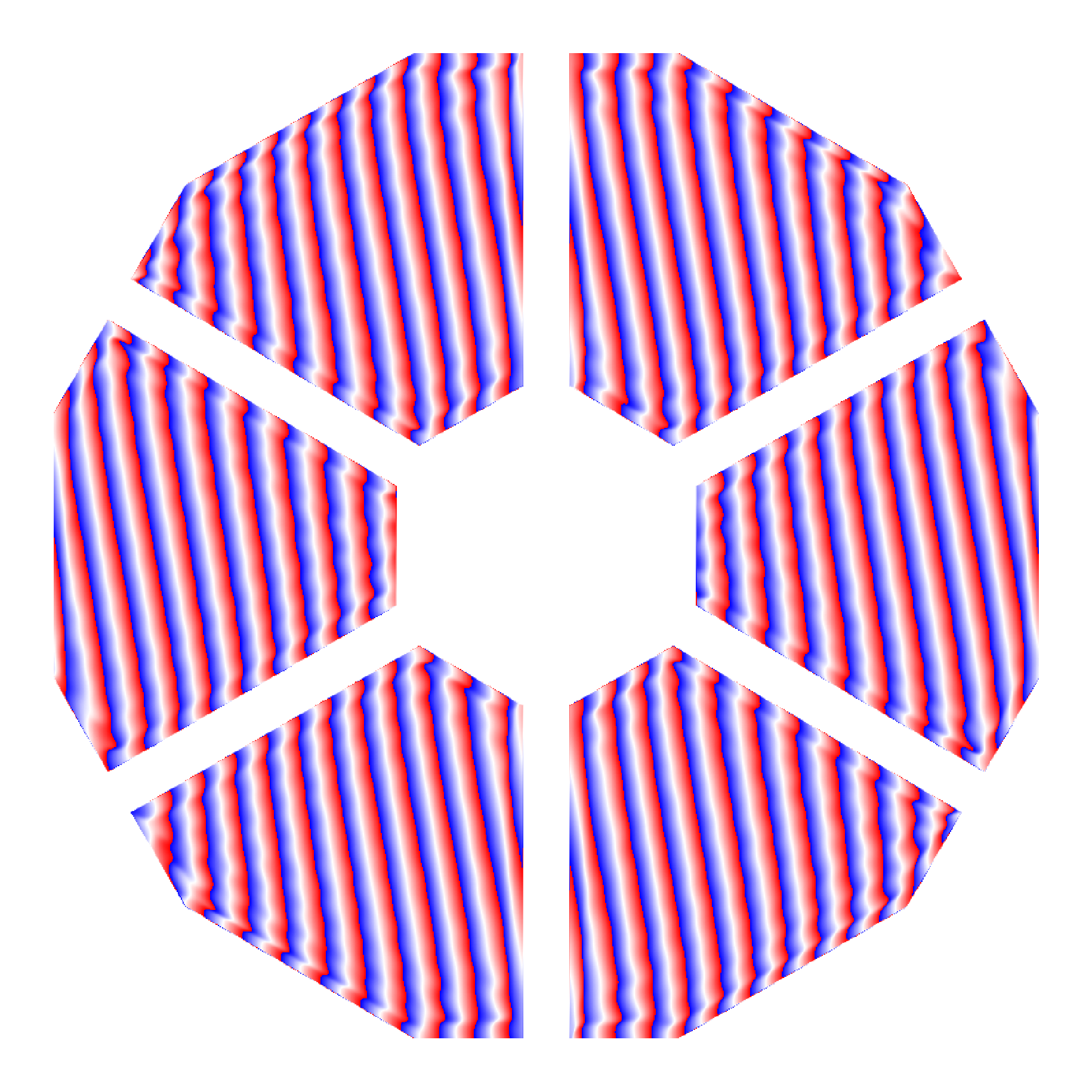}
    \end{subfigure}
    ~
    \begin{subfigure}[c]{0.3\textwidth}
        \centering
        \includegraphics[width=\linewidth]{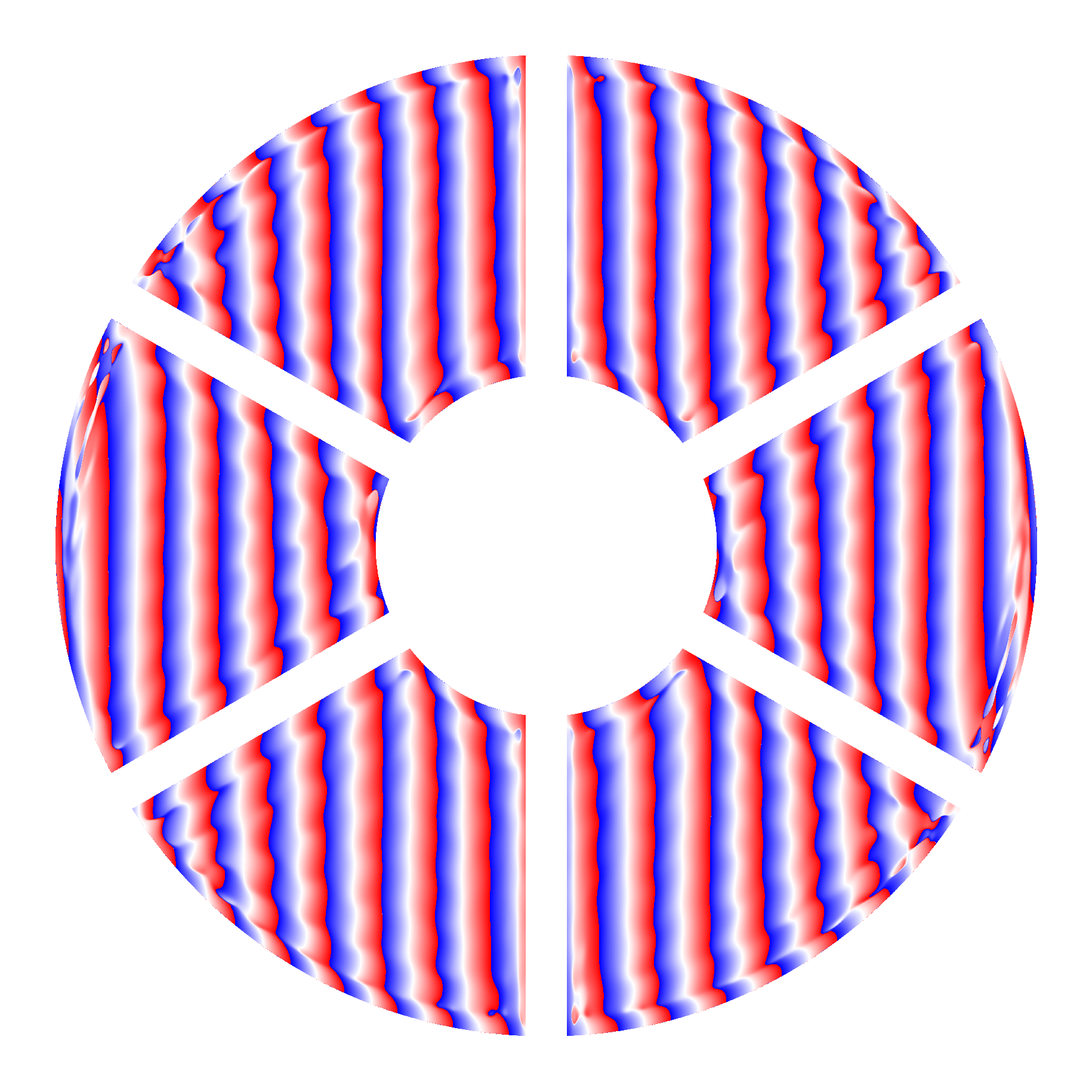}
    \end{subfigure}
    \\ % vAPP figures
    \begin{subfigure}[c]{0.3\textwidth}
        \centering
        \includegraphics[width=\linewidth]{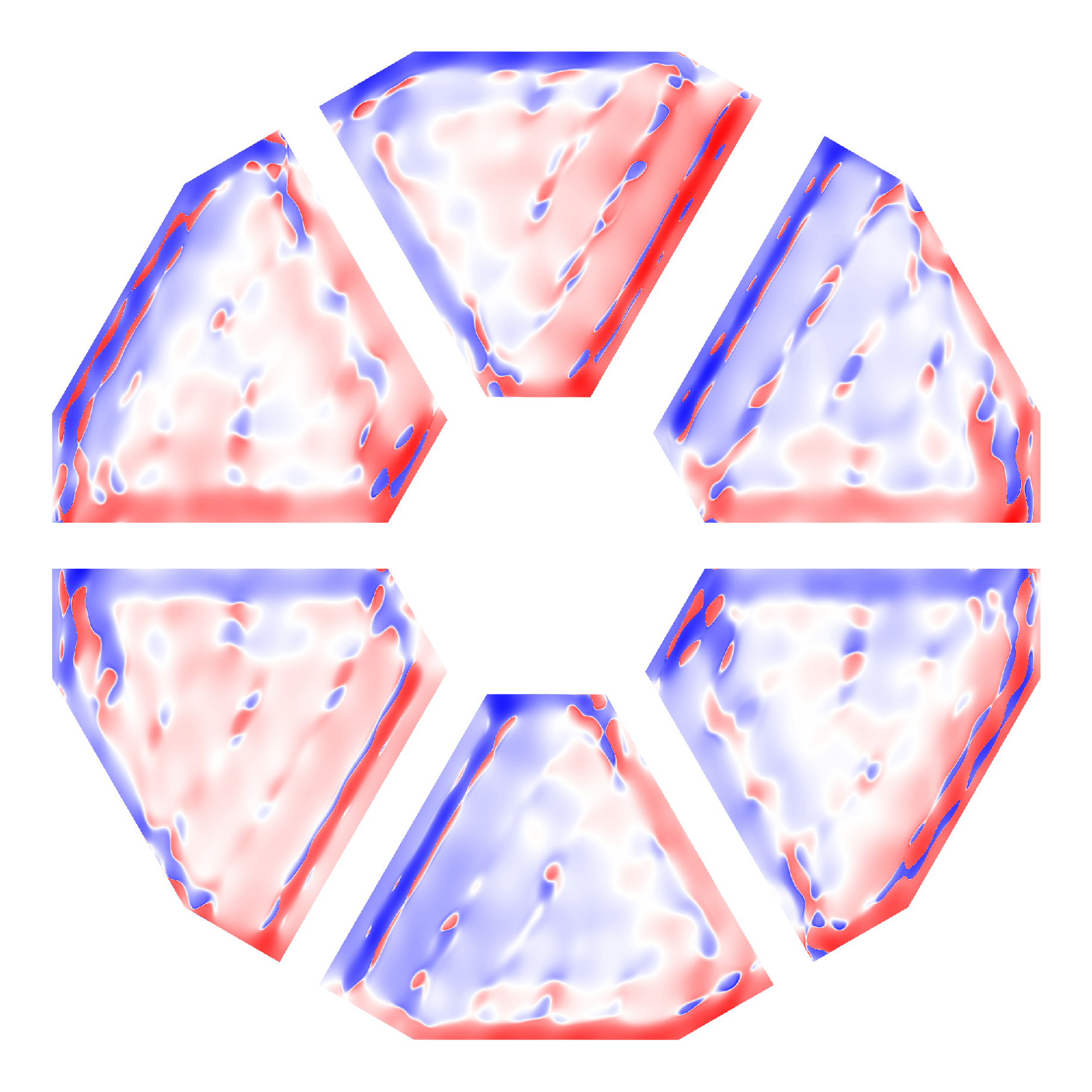}
    \end{subfigure}%
    ~ 
    \begin{subfigure}[c]{0.3\textwidth}
        \centering
        \includegraphics[width=\linewidth]{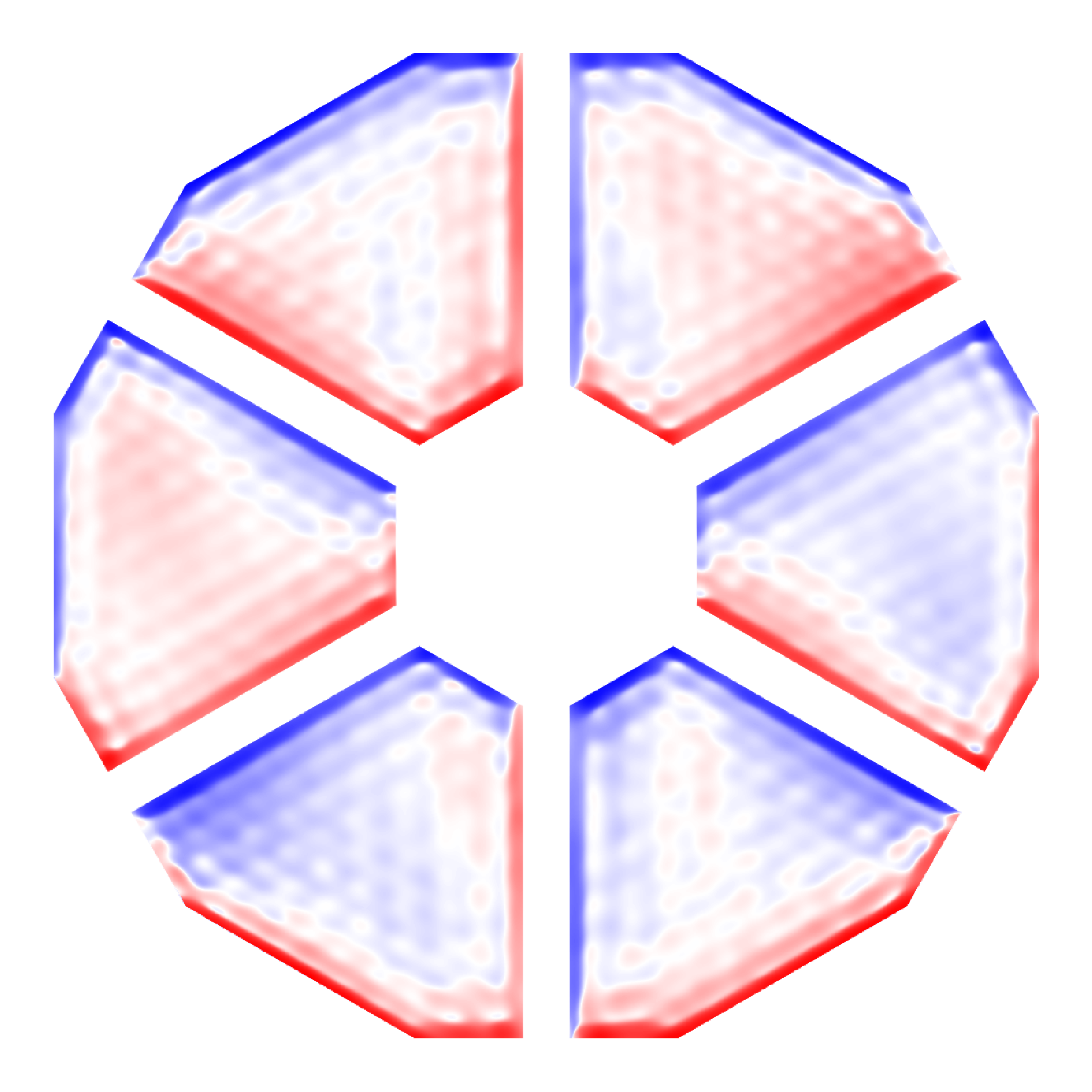}
    \end{subfigure}
    ~
    \begin{subfigure}[c]{0.3\textwidth}
        \centering
        \includegraphics[width=\linewidth]{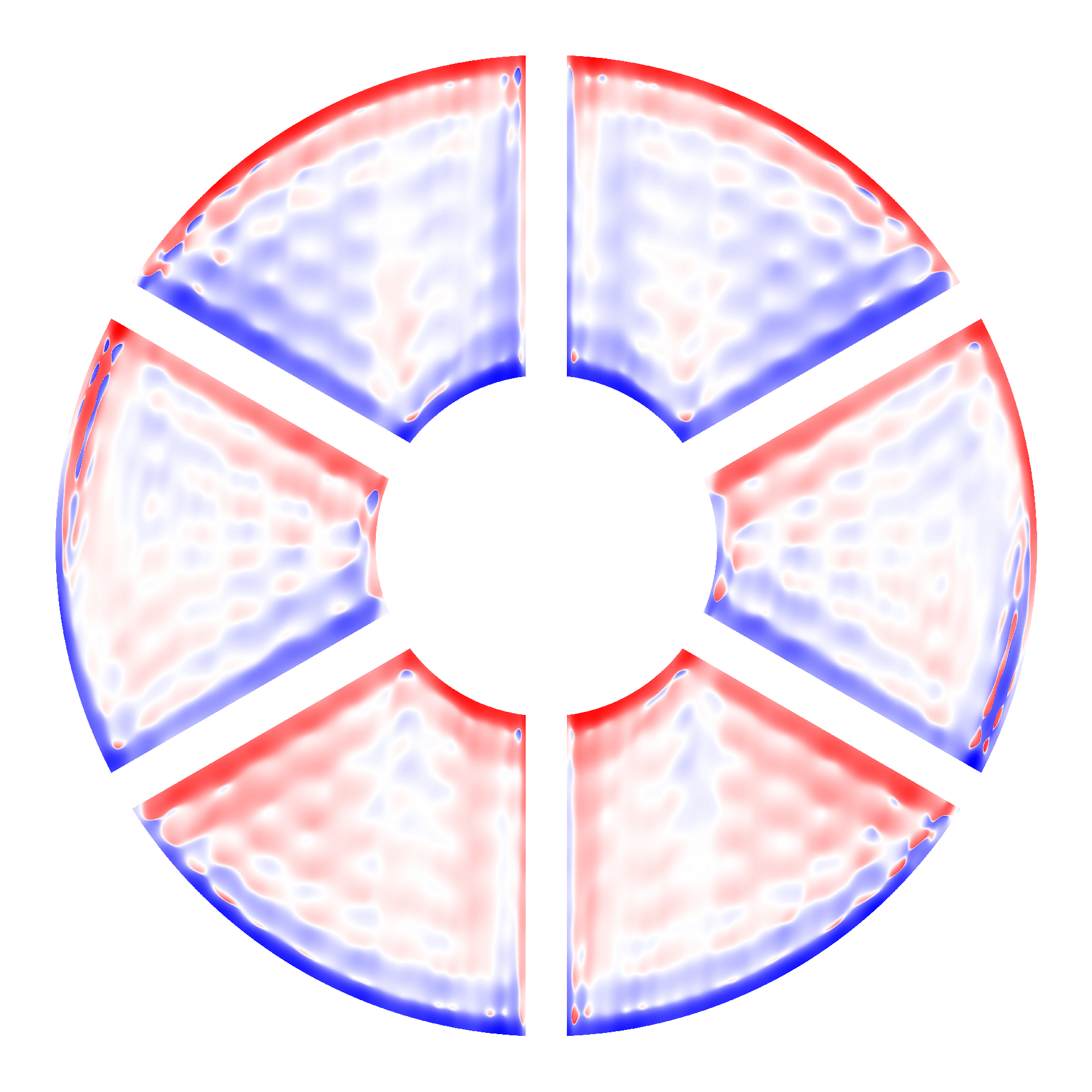}
    \end{subfigure}
    \\ % PSF figures
    \begin{subfigure}[c]{0.3\textwidth}
        \centering
        \includegraphics[width=\linewidth]{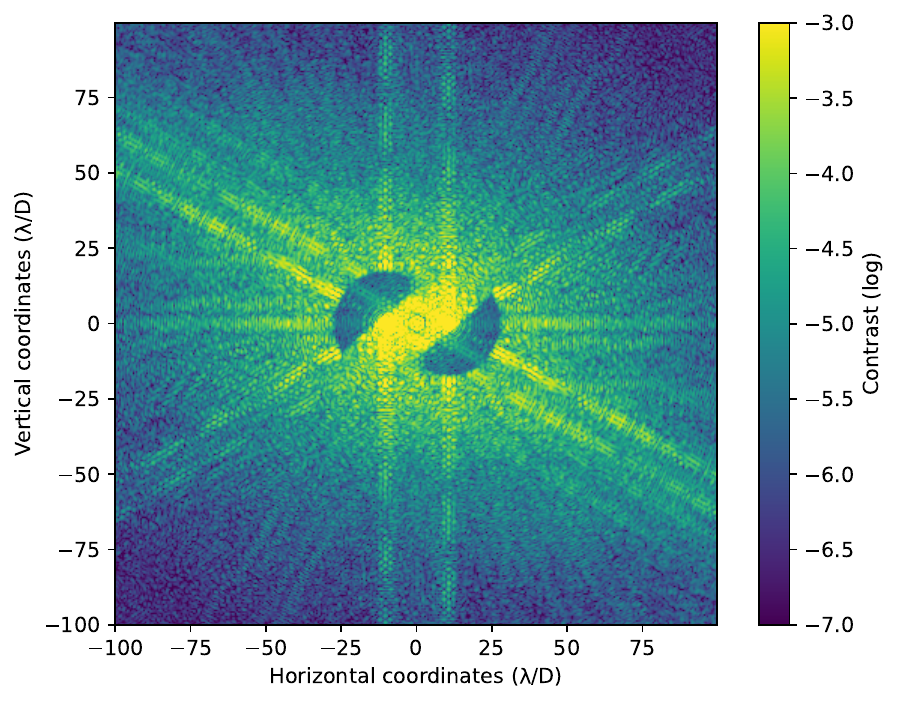}
        \caption{METIS APP-IMG}
    \end{subfigure}%
    ~ 
    \begin{subfigure}[c]{0.3\textwidth}
        \centering
        \includegraphics[width=\linewidth]{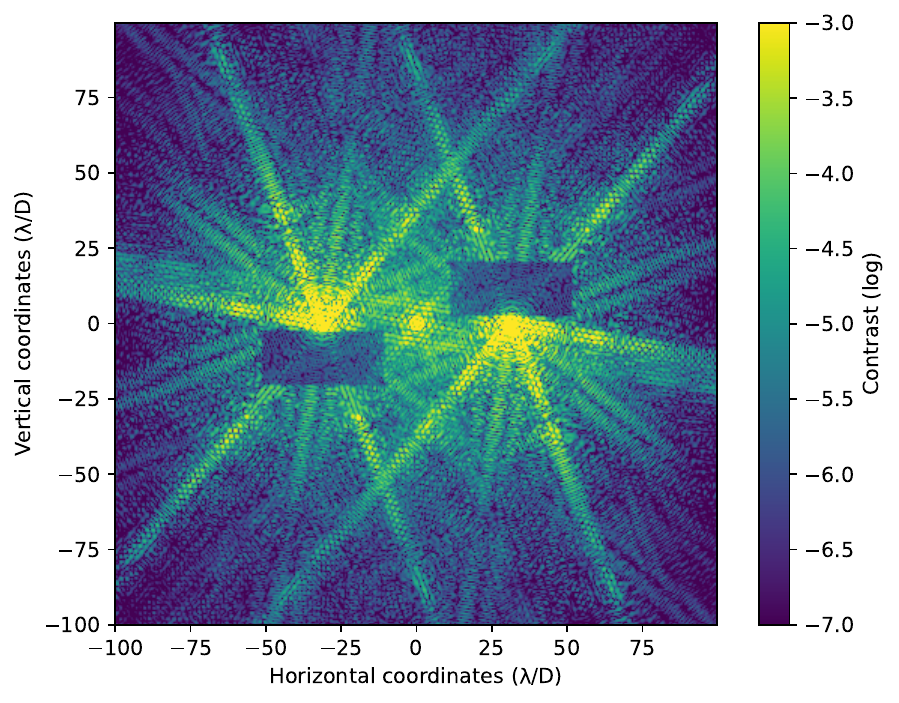}
        \caption{METIS APP-LMS}
    \end{subfigure}
    ~
    \begin{subfigure}[c]{0.3\textwidth}
        \centering
        \includegraphics[width=\linewidth]{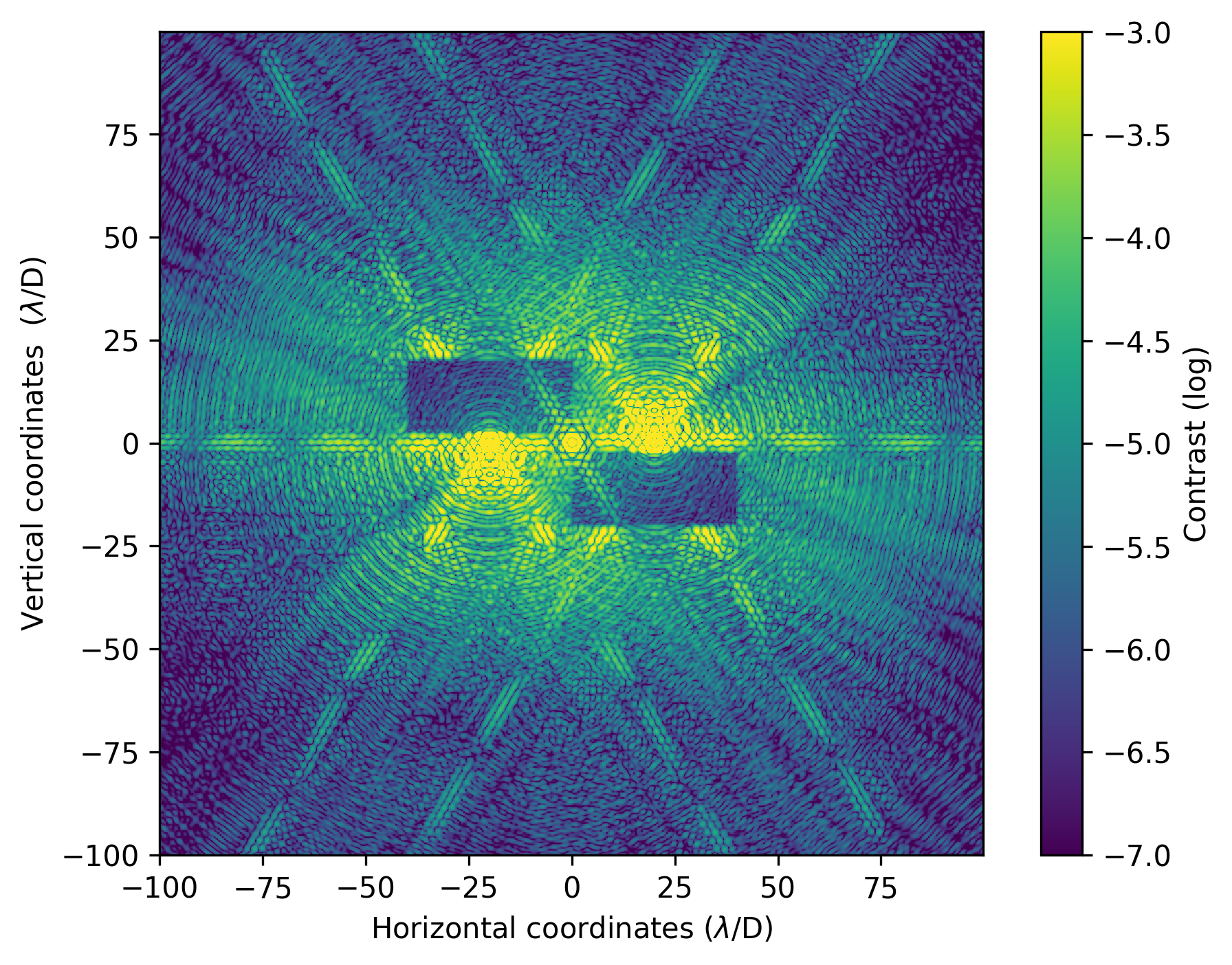}
        \caption{MICADO \acs{vAPP}}
    \end{subfigure}
    \caption{This figure provides an overview of the designs of the three \acs{vAPP} in this proceeding. The left column illustrates the design of the \acs{vAPP} for the METIS LM imager, the middle column that for the METIS LM spectrograph and the right column the design for MICADO. The top row shows the full phase pattern. The middle row shows it with the grating pattern subtracted. The bottom row shows the resulting PSFs at the same spatial and color scale.}\label{fig:APP_designs}
\end{figure*}

We will briefly describe the design commonalities and differences in the design of the \acs{vAPP}s for both \acs{METIS} and \acs{MICADO}. A tabulated overview of the specifications is given in Table~\ref{tab:APP_specs}. The phase pattern designs and their corresponding PSFs are shown in Fig.~\ref{fig:APP_designs}.

\begin{table}[tb]
    \centering
    \caption{Overview of the specifications of the three \acs{gvAPP}.}
    \small
    % \begin{tabular}{p{0.5cm}|p{3cm}|c|c|c}
    \begin{tabular}{p{0.5cm}p{3cm}ccc}
        \toprule
        % & & \multicolumn{2}{c}{METIS} & MICADO \\
        & & \textbf{METIS APP-IMG} & \textbf{METIS APP-LMS} & \textbf{MICADO} \\ \midrule %\hline
        & Wavelength range & 2.9 - 5.3 \textmu m & 2.9 - 5.3 \textmu m & 1.1 - 2.4 \textmu m \\
        \parbox[t]{2mm}{\multirow{7}{*}{\rotatebox[origin=c]{90}{PSF design}}} & Dark zone shape & Semi-circle & Rectangular & Rectangular \\
        & IWA & 2.5$\lambda/D$ & 2.5$\lambda/D$ & 2.6$\lambda/D$ \\
        & OWA & 17$\lambda/D$ & $(30 \times 45)\lambda/D$ & $(20 \times 40)\lambda/D$ \\
        & Separation between conjugated PSFs & 20$\lambda/D$ & 60$\lambda/D$ & 40$\lambda/D$ \\
        & Contrast & 10\textsuperscript{-3.5} at IWA & 10\textsuperscript{-3.5} at IWA & 10\textsuperscript{-3.0} at IWA \\
        & & 10\textsuperscript{-5.5} at OWA & 10\textsuperscript{-5.5} at OWA & 10\textsuperscript{-6.0} at OWA \\
        & Strehl: Relative to unapodized pattern & 60\% & 72\% & 70\% \\ \midrule %\hline
        \parbox[t]{2mm}{\multirow{8}{*}{\rotatebox[origin=c]{90}{Optical assembly}}} & Substrate 1 & CaF2 & CaF2 & Infrasil 302 \\
        & \phantom{000}Dimensions & 52 mm $\times$ 2.5 mm & 30 mm $\times$ 4.0 mm & 100 mm $\times$ 8 mm \\
        & \phantom{000}Coating(s) & Chrome mask, AR coating & Chrome mask, AR coating & Chrome mask, AR coating \\
        & Substrate 2 & CaF\textsubscript{2} & CaF\textsubscript{2} & Infrasil 302 \\
        & \phantom{000}Dimensions & 48 mm $\times$ 2.5 mm & 24 mm $\times$ 2.5 mm & 90 mm $\times$ 3.5 mm \\
        & \phantom{000}Coating(s) & Phase pattern & Phase pattern & Phase pattern \\
        & Alignment feature & Flat side & Flat side & V-notch \\
        & Wedge & 0.3\degree\ & 0.3\degree\ & None \\ \midrule %\hline
        \multirow{4}{*}{\rotatebox[origin=c]{90}{\parbox[b]{1.5cm}{\centering Pattern\\\& Recipe}}} & Pattern diameter & 43.73~mm & 19.27~mm & 74.41~mm \\
        & Pattern Pixel size & 10 \textmu m & 10 \textmu m & 14 \textmu m \\
        & Recipe design & 3TR & 3TR & 3TR \\
        & Leakage & $<2$\% & $<2$\% & $<1$\% \\ \midrule %\hline
        & Other & \parbox[c]{2.5cm}{\vspace{0.1cm}FP-WFS through ALF\cite{Orban2024,Bos2019,Bos2021}\vspace{0.1cm}}  & - & - \\ \bottomrule%\hline
    \end{tabular}
    \label{tab:APP_specs}
\end{table}

First, METIS will feature two different \acs{vAPP}s: One for the \textit{LM}-imager\cite{METIS_IMG2024} (APP-IMG) and one for the \textit{LM}-spectrograph (APP-LMS). Both optics consist of a CaF\textsubscript{2} (Calcium Fluoride) stack, with an anti-reflection coating on the outwards facing surfaces. The optically active \acs{LCP} coating, NOA-61 optical adhesive, and an amplitude mask are located between the two plates. The APP-IMG features a D-shaped dark zone, with an \ac{IWA} of 2.5$\lambda/D$ and an \ac{OWA} of 17$\lambda/D$, while the APP-LMS features a rectangular dark zone of about 0.58 by 0.90 arcsec (approximately, 30 by 45 $\lambda/D$ in \textit{L}-band) closely matching the dimensions of the \textit{LM}-spectrograph slicer.

Next, the MICADO \acs{vAPP} is designed with a rectangular dark zone starting at an \ac{IWA} of $2.6\lambda/D$ and an \ac{OWA} of $20\lambda/D$. In contrast to the smaller sized substrates of METIS -- 50~mm diameter for the APP-IMG and 30~mm for the APP-LMS, MICADO features a substantially larger cold pupil at 82~mm. Therefore the pupil optics have a diameter of 100~mm. The assembly of the stack is otherwise similar.

Because the phase pattern coating is fragile, a second substrate is glued on top of the coated substrate to provide protection. For all three \acs{vAPP}s, this second substrate features a chrome mask that is slightly undersized compared to the nominal optical pupil of either instrument. In this way, the fundamental design assumption of uniform illumination is closely met in practice.

Another important design aspect is the relative orientation of the phase pattern compared to the instrument optical pupil and the optic projects the PSF onto the detector. Some systematic effects can be difficult to deal with after the design is final\cite{Kenworthy2026}. One example are electronic read-out ghosts. These appear when bright parts in the image, such as the PSF cores, are read out by the CCD detector and they present themselves as a negatively valued imprint in nearby detector channels\cite{Finger2008}. Although this effect is well known and has been reduced to low levels in modern detectors, their impact can easily be avoided by making sure that the bright PSF cores and the detector rows do not align. Furthermore, the direction of PSF diffraction spikes should ideally pass through as little of the dark zones as possible. Aligning these different constraints requires a careful inspection of the optomechanical design and the (coordinate) transformations and between the pupil plane and the image plane.

%% file: 02_manufacturing.tex
\section{Summary of the manufacturing process}\label{sec:manufacturing}

\begin{figure}[b!]
    \centering
    \includegraphics[width=0.8\linewidth]{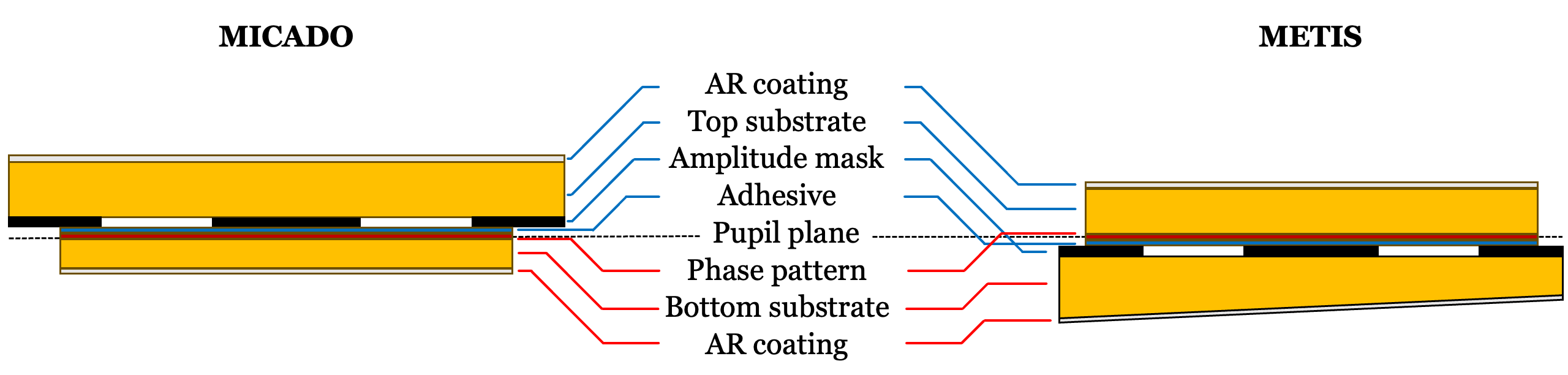}
    \caption{The general anatomy of a \acs{vAPP} consists of two substrates, which feature an \ac{AR} coating on the outside surfaces. The two substrates are glued together using an adhesive to protect the fragile \acs{LCP} based pupil pattern. Also, an amplitude mask is added there to ensure even illumination of the pupil.}
    \label{fig:APP_anatomy}
\end{figure}

During the design phase of the \acs{vAPP}, the optical and mechanical interfaces guide the anatomy of the optical assembly, shown in Fig.~\ref{fig:APP_anatomy}. As these are pupil-plane coronagraphs, the phase pattern and amplitude mask should be as close to the optical pupil as possible. The larger of the two substrates is typically used to mechanically position the optic in its mount. A flat edge or V-notch can help to fix the clocking angle. Lateral pupil alignment is mechanically guided by the position tolerances of the optic in the mount and optically by the absolute positioning tolerance of the amplitude mask -- or phase pattern if this is deposited on the larger of the two substrates. \acl{AR} (\acs{AR}) coatings and potentially a wedge in the lower substrate reduce the risks of unwanted ghost images.

The manufacturing process naturally starts with the procurement of the optical substrates, followed by an application of the \acs{AR}-coatings on what will end up being the outer surfaces of the assembly. For METIS, Calcium Fluoride (CaF\textsubscript{2}) substrates are used in tandem with an AR coating with less than 2\% reflection between 2.0 and 5.3~\textmu m. In MICADO, Infrasil substrates and an AR coating with $R_\mathrm{max} < 1\%$ between 1.15 and 2.35~\textmu m form the basis for the \acs{vAPP}. Then, a chrome based mask was photolithographically printed on the larger substrates. 

The smaller non-wedged substrates are used for the pupil phase pattern deposition. The process is based on liquid-crystal technology. A thin liquid-crystal based photo-alignment layer, the \ac{LPP} layer, is spin-coated on the substrate. A direct-write system then inscribes the desired orientation into the \acs{LPP} layer with a UV laser ($\lambda = 355$~nm) that can change the direction of the linearly polarized light it emits \cite{Miskiewicz2014}. Next, multiple layers of \acl{LCP} (\acs{LCP}) material are spin-coated on top and subsequently cured in place through UV exposure. The liquid crystal polymers self-align based on the underlying polymer orientation\cite{Schadt1995b}. Highly efficient performance can be obtained, approaching near perfect half-wave retardance over a wide wavelength range, by designing and optimizing a recipe consisting of three \ac{MTR} layers\cite{Komanduri2013}. The optical twist and thickness are varied along the stack according to this predetermined recipe. The thickness of the complete \acs{LCP} stack strongly depends on the birefringence of the specific \acs{LCP} material that is used and the wavelength range for which it is optimized. For the MICADO optic, the total stack is about 11~\textmu m thick and for both the METIS \acs{vAPP}s the thickness reaches up to 35~\textmu m. It is not possible to coat such thick layers at high quality with only a small number of spin-coating runs. Therefore, the three main twist retarder layers are divided into thinner sublayers. By measuring the coating progression after each sublayer deposition with an optical and near-infrared spectroscopic ellipsometer, it is possible to retrieve up-to-date information about the production variations and adjust the spin-coater settings to return to the optimum recipe. Specifically for this purpose, the MICADO \acs{vAPP} phase pattern featured a central test pattern in the part that will later be covered by the amplitude mask.

When the phase pattern is finalized, the last remaining step is to align the two substrates and laminate them together with an optical adhesive suitable for use in a cryogenic environment. Alignment features are added to the chrome mask and the phase pattern. These consist of a microscopic dot (diameter: 100~\textmu m) and a corresponding circle with the same radius. They are located outside the clear aperture, near the edge of the substrates. With the use of a polarization microscope, the two substrates can be very accurately aligned, held in place and then the adhesive can be fully cured. For the adhesive, Norland Optical Adhesive (NOA-61) is used.

%% file: 03_performance_indicators.tex
\section{Performance indicators}\label{sec:leakage}

The self-aligning property of the \acs{LCP} material has as a very useful consequence that it is possible to separate testing of the phase pattern from that of verifying the recipe (i.e. the polarization leakage as a function of wavelength). In this way, the design of the phase pattern can be tested on a sample with a much simplified \acs{LCP} coating and the resulting PSF can be characterized using smaller optics and more common wavelengths. The prototype pattern designs of the METIS and MICADO \acs{vAPP}s were presented at the previous iteration of this conference\cite{Doelman2024}. There, excellent agreement was obtained between simulation and the measured \acs{PSF}. 

The development, characterization and verification of the coating recipe can be done in two ways. One approach is to deposit the coating on a grating pattern to create a polarization grating\cite{Oh2008}. Most light incident upon this grating should then be diffracted into the first diffraction orders. Any deviation from perfect half wave retardance will cause a fraction of the light to propagate towards the zero'th order -- straight through. This is the polarization leakage fraction. Spatial filtering of the light beam then allows for easy measurement of the leakage fraction with a spectrometer. If the same measurement is performed on a part of the phase pattern that does not feature the grating, then the overall transmission of the optic is obtained. 

A second method requires the use of a spectroscopic ellipsometer. By measuring the full Mueller matrix of the coating, and in particular that of the $M_{4,4}$ matrix element, the polarization leakage fraction can be estimated. This matrix element describes the transformation of circular polarization by the retarder. For a perfect half wave retarder, $M_{4,4}=-1$. Any deviation from this will lead to incomplete polarization conversion and therefore leakage. For an idealized half-wave retarder, the leakage fraction is given by
\begin{equation}
    L = \frac{1 + M_{4,4}}{2}.
\end{equation}
To first order, this value is equivalent to the fraction of light measured by the first method.

\begin{figure}[tb!]
    \centering
    \includegraphics[width=\linewidth]{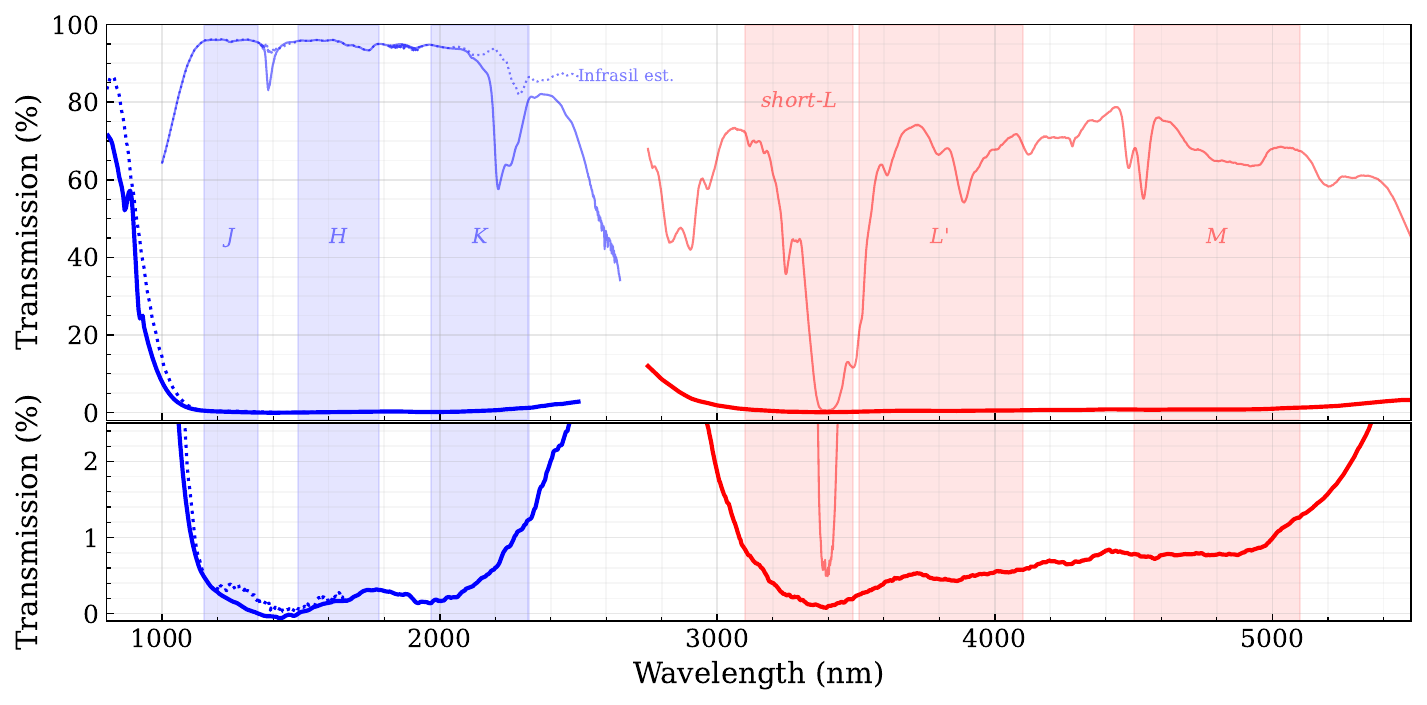}
    \caption{The overall transmission (thin lines) and polarization leakage (thicker lines) of the METIS (red) and MICADO (blue) \acs{vAPP} coating recipe. The dotted line on the left is the polarization leakage derived from ellipsometer measurements. For MICADO, the measurement was done on test sample of optical grade fused silica, explaining the absorption near 2200~nm. An extrapolation to Infrasil 302 is also presented. For METIS, a measurement of the final CaF\textsubscript{2} APP-IMG optic is shown. Exceptionally low polarization leakage is demonstrated over a very wide passband for both METIS and MICADO, as illustrated in the lower panel.}
    \label{fig:transmission_and_leakage}
\end{figure}
% MICADO: 1 octave of bandwidth below 1% leakage. Fractional bandwidth: 0.688
% METIS: 0.7 octave of bandwidth below 1% leakge. Fractional bandwidth: 0.476
Figure~\ref{fig:transmission_and_leakage} presents the end result of the recipe development process. Exceptionally low leakage is demonstrated over a wide and useful wavelength range, where we have found an average of 0.6\% leakage for the METIS recipe and only 0.3\% for MICADO. In fact, for MICADO the leakage is below 1\% for more than an octave in bandwidth. The polarization leakage of the recipes were determined through the polarization grating method. For MICADO, the polarization leakage measurement was also verified up to 1690~nm by measurements with a J.A. Woollam RC2 ellipsometer. Unfortunately, no such verification was possible for the mid-infrared wavelengths of the METIS \acs{vAPP}. The overall transmission was performed on a representative test sample for MICADO (laminated optical grade fused silica substrates with all coatings) and on the final APP-IMG optic for METIS. The major absorption feature seen at 3400~nm is a result of the molecular vibration modes of the polymers in both the adhesive and liquid crystal materials. A similar absorption feature was found for the ERIS \acs{vAPP} too\cite{Kenworthy2026}.

Although these independent tests are useful to gain confidence in the production process, compliance can only definitively be demonstrated on the final optic. However, in the era of ELT instrumentation, such tests quickly increase in complexity. In particular the need for diffraction limited performance, low (thermal) background, a detector that works at the desired wavelengths and the high dynamic range that is needed to validate the contrast levels mean that suitable test benches are full instruments by themselves\cite{Boehle2021}. Therefore, of the three optics discussed here, only the APP-LMS will be tested in the lab, and only then at room temperature. The VODCA test bench\cite{VODCA2019} is already being used to characterize the vortex coronagraphs of METIS and is therefore an excellent test bench to characterize the APP-LMS as well. Once the APP-LMS production has finished, we plan to perform measurements there.

%% file: 04_defects.tex
\section{Challenges in the production process}\label{sec:defects}
% \epigraph{\textit{No plan survives first contact with the enemy.}}{Helmuth von Moltke the Elder (paraphrased)}

The idealized worlds of theory and simulation come to meet with the challenges of reality during manufacturing. In this section, we will discuss some of the widely varying difficulties we have encountered during the various production stages of these optics, from the writing of the phase pattern to the lamination process. We also discuss how we have subsequently dealt with them and what their impact on the optical performance is.

\subsection{Observed defect modes during production}
\begin{figure}
    \centering
    \begin{minipage}[t]{.32\textwidth}
        \includegraphics[width=\linewidth]{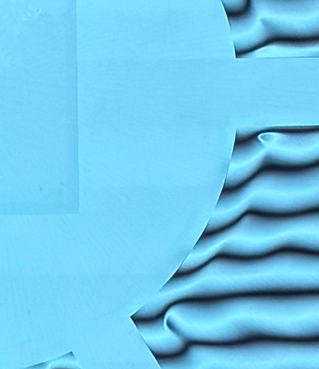}
    \end{minipage}
    \begin{minipage}[t]{.66\textwidth}
        \includegraphics[width=\linewidth]{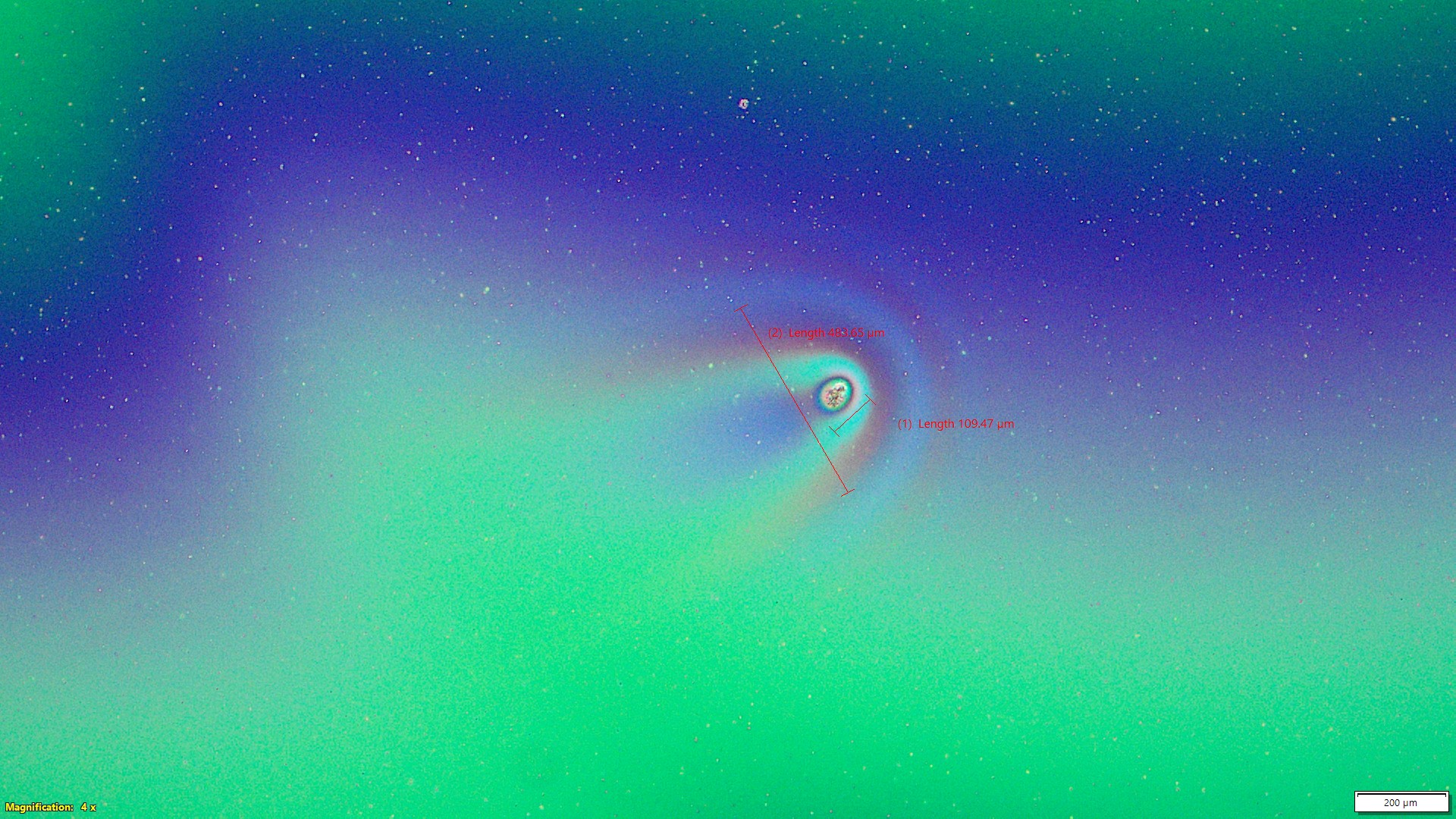}
    \end{minipage}
    \caption{Examples of some of the observed defects. The left picture shows the shadow effect seen after printing the phase pattern. This presents as the horizontal lines resembling a trailing shadow after a fast change in the pattern. The right picture shows an example of the bowshock defect, where some crystallized \acs{LCP} material (diameter: 100~\textmu m) influences the coating distribution around it. The tail points radially outward and is approximately 0.5~mm long.}\label{fig:defects_examples}
\end{figure}

The METIS APP-IMG was the first optic intended for mid-infrared observations that was manufactured and delivered by \acl{CLJ} Due to the long operational wavelength, the complete coating stack required many sublayers, making the production process complex and time consuming. An extensive research and development phase was necessary to improve production quality before the manufacturing of the real \acs{vAPP} could be started. Even then, careful inspections of the optic during the production revealed various coating defects. For many of these, their origins are now well understood, allowing us to reduce the occurrence frequency and minimize their impact. We list some of the more common localized defects in the coating process below. Two examples are presented in Fig.~\ref{fig:defects_examples}.

\begin{itemize}
    \item Calibration errors: Nonlinearities in the system response and other systematic effects cause discrepancies between the desired and obtained fast axis orientation, known as pattern writing errors. Often these appear as undesired phase jumps in the produced pattern.
    % \item Voltage calibration error: Nonlinearities and boundary errors between the voltage and polarization angle transmitted by the Pockels cell of the direct write machine result in pattern writing errors, such as a jump of the written fast axis angle as a function of the applied voltage. The amplitude of these effect can be partially attributed to environmental factors, such as the room temperature.
    \item \acs{LPP} spot defect: A localized region with a different fast axis angle than designed.
    \item \acs{LPP} shadow: A line of decreasing intensity that occurs after a large phase angle jump in the writing direction. It is a function of the amplitude of the phase jump and the velocity of the translation stage.
    \item Particle defects: These were observed as either contamination -- particles originating from elsewhere that land on the coating layer during the production process -- or as crystallized \acs{LCP} material that formed in the coating reservoir. 
    \item Bowshock defects: When particle defects remain on a lower coating layer and one or multiple layers are spin-coated on top, a region of higher and lower coating density appears near the original defects. This has the shape of a shockwave towards the center of the optic.
    \item Coating non-uniformity: Radial misalignment between the rotation axis of the optic in the spin-coating machine and the deposited \acs{LCP} droplet can result in small nonuniformities of the layer thickness. While this thickness variation is small for individual sub-layers, they can add up over multiple layers. Visually, this can be seen as a color variation, when inspecting the optic under cross-polarizers.
    \item Edge bead: Surface tension characteristics of the \acs{LCP} will have the unintended result that slightly more coating material collects near the edge of the substrate\cite{Yan2018,Yan2021}. For one sublayer, this effect is minimal, but with many layers this effect accumulates.
\end{itemize}

\begin{figure}
    \centering
    \includegraphics[width=.85\linewidth]{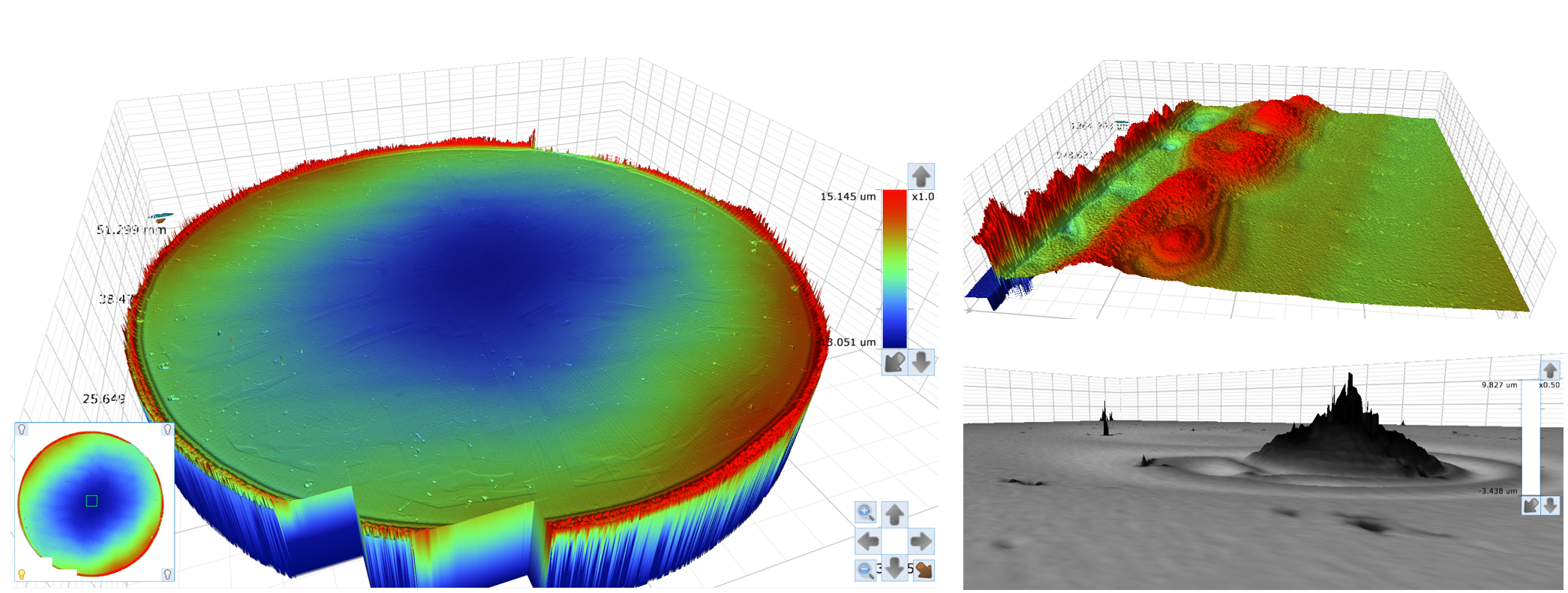}
    \caption{Using a Bruker NPFLEX optical profilometer, we were able to measure the surface profile of the coated surface. The thickness variation across the substrate and around local defects could then be measured. This measurement confirms some of our suspicions about observed defects and allows us to better estimate the performance impacts on the final assembled optic. \textit{Left}: Stitched profile of the complete substrate. \textit{Top right}: Surface view of the edge bead. \textit{Bottom right}: Three dimensional view of a local defect with a surrounding bowshock in the coating material.}
    \label{fig:bruker}
\end{figure}

Through classification, detailed inspections and experiments, we have determined the origins of these defects and in most cases reduced their impact. The first three of these defects can be detected early in the production process, through calibration samples and visual inspections\cite{Miskiewicz2014,landman2026}. If the result is not acceptable, the \acs{LPP} layer can be removed and substrate can be reused. The latter three defect types affect the (local) thickness of the coating layer and can be modified through changes in the coating environment or by varying the spin coater parameters. The thickness variations can be seen as color variations under cross-polarizers and these change the retardance properties when these propagate through the main twist retarder layer that the coating sublayer is part of. Recently, we have been able to confirm the hypothesized surface profile variations using a Bruker NPFLEX optical profilometer. A rejected production sample of the METIS APP-IMG was measured. A clear bowl shape with a strong edge bead near the substrate edge was observed. Also, an inspection of bowshock defects clearly show varying levels of accumulation around the central particle. This is illustrated in Fig.~\ref{fig:bruker}.

The final step in the assembly of the \acs{vAPP} is the lamination of the coated substrate with the protective substrate containing the chrome mask. Both the METIS and MICADO \acs{vAPP} are part of the cold optics instrument and will be cooled down to 65~K and 82~K during operations, respectively. Such a large temperature differential between assembly and operational conditions can create stress from differences in the \ac{CTE} that the optic and the adhesive must be able to handle. We have selected Norland Optical Adhesive 61 (NOA-61) as the preferred adhesive. It has been already been used for laminated optics in cryogenic instruments\cite{Otten2017}, it has been qualified for space usage, and it has excellent optical characteristics in terms of transmission and refractive index. To validate its usability, we have cryogenically cycled multiple test samples of laminated CaF\textsubscript{2}, and later also of laminated Fused Silica substrates. The dimensions of these test samples were representative of the intended application. One qualification sample of the METIS APP-IMG, one not intended for use in the final instrument, was also used for a cryogenic thermal cycle.
% Aside from optical transmission requirements, such a large temperature differential between assembly and operational conditions can create stress from differences in the \ac{CTE} that the optic and the adhesive must be able to handle. For this reason, we selected Norland Optical Adhesive 61 (NOA-61). Already, it has been used in cryogenic instruments\cite{Otten2017} and it has been qualified for space usage. To validate its usability, we have cryogenically cycled multiple test samples of laminated CaF\textsubscript{2}, and later also of laminated Fused Silica substrates. The dimensions of these test samples were representative of the intended application. One qualification sample of the METIS APP-IMG, one not intended for use in the final instrument, was also used for a cryogenic thermal cycle.

For the laminated CaF\textsubscript{2} substrates minor damage was found. Tiny fragments of the glass broke of near the glass-to-glue interface. Repeated thermal cycles showed that these were a release of excess stress that occurred during the first cooldown. The stress induced by the thermal shrinkage differences must have exceeded the fracture point of the glass at some critical points. Notably, the damage did not increase after the initial cooldown and only appeared near the outer diameter of the glass, without propagating towards the clear aperture. Furthermore, the severity of the damage did not scale with the size of the optic, as smaller samples only 25~mm in diameter showed more fragmentation than a different test sample that was 50~mm in diameter, suggesting that the adhesion thickness and the quality of interface might play an important role. An adjustment to the adhesion procedure to carefully remove any excess adhesive from the assembled optic and the extra validation that the qualification piece survived a thermal cycle intact, provides sufficient confidence in the suitability of CaF\textsubscript{2} as the substrate material for the METIS \acs{vAPP}s.

For MICADO, Infrasil 302, a type of Fused Silica optimized for the infrared, is used for the substrates. The laminated substrates have a diameter greater than recommended by Norland\cite{Norland1995}. Nonetheless, no visual indications of damage were observed after two cycles between room temperature and 90~K. A cleaner adhesive interface and higher fracture toughness compared to CaF\textsubscript{2} likely explain the result, which is certainly advantageous with respect to the large diameter of the optics required for the MICADO \acs{vAPP}.

Of more concern for this optic is the wavefront quality. During preparations for the production of the final optics a test sample was manufactured with a fully representative coating and adhesive. A significant portion of the phase pattern was left blank to allow us to measure the transmitted wavefront error after. Measurements using a Zygo laser interferometer (at 632~nm), a Phasics Kaleo MultiWAVE (at 1550~nm) and a visual inspection with a Sodium light (at 589~nm) all revealed multiple waves of wavefront error, primarily contained in a defocus term. For a final optic, such a large wavefront error in transmission would be unacceptable.

The substrates used for this test sample were of less optical quality than the optics intended for the final product, but could not alone explain the measurement. The surface profile of the liquid crystal coating also would not provide a sufficiently large error. In fact, the 
wavefront error of the test sample increased from an \ac{RMS} value of 52~nm before the coating, to 162~nm after the \acs{LCP} coating was applied, to 753~nm after the lamination was performed. This measurement has taught us that even this last step in the assembly cannot be taken lightly. Since then, we have identified several steps in the lamination procedure that should be investigated and can be improved upon, such as the adhesive deposition method, the method to spread out the adhesive and the UV exposure parameters to cure the adhesive\cite{Norland1993,Norland1995,Xiong2026}. A deeper investigation and multiple experiments are now planned to determine the best approach to laminating these large substrates.

\subsection{Performance analysis}
% Do the defects impact the performance?

Given the observed defect modes, we now attempt to analyze the impact of most of these on the optical performance of the METIS APP-IMG and the MICADO \acs{vAPP}. The main performance metric is the raw contrast level as a function of radial distance from the peak of one of the \acs{PSF} copies. In all cases, we normalize the brightest part of the \acs{PSF} to an intensity level of unity and assess the contrast levels as an azimuthal average in the dark zone, away from the PSF core. The analysis was done in Python and made heavy use of \texttt{HCIPy}\cite{por2018hcipy} for simulation of the \acs{vAPP}s.

\begin{table}[b]
    \centering
    \caption{This table specifies the type of defects that were included in the performance simulation. The first column denotes the defect type, the second column what optical property is affected. The third column provides some specifics about the defect based on what was seen for the manufactured METIS APP-IMG. The last column provides an indication of how the defect impacts the PSF.}
    \small
    \begin{tabular}{p{3.5cm} p{3.5cm} p{3.3cm} p{5cm}}
    \toprule
    Defect type & Affected property & Value & Result on PSF\\
    \midrule
    \acs{LPP} spot defect & Fast-axis orientation & $2\times\text{\diameter}60$~\textmu m & Negligible change of PSF shape.\\
    Fast-axis angle write error & Phase pattern & 3\% & Decreased contrast, decreased Strehl ratio, half-order diffraction.\\
    Polarization leakage & Retardance & 1.5\% & Light in leakage PSF.\\
    \acs{LCP} inclusions and bowshocks & Amplitude (core), retardance, fast-axis angle & $6\times$ (Core:\diameter\ 40-120~\textmu m.\newline Bowshock: $\leq2.5$~mm.) & Light in leakage PSF\\
    Misalignment phase pattern and chrome mask & Amplitude & 10~\textmu m & Light in leakage PSF\\
    Air bubble in adhesive & Amplitude (edge), phase (void) & $1\times250$~\textmu m & Minor change in PSF shape and contrast.\\
    Coating thickness variation & Retardance & Peak-to-valley variation: 10\% & Light in leakage PSF\\
    \bottomrule
    \end{tabular}
    \label{tab:sim_defects_spec}
\end{table}

% \begin{figure}[tb!]
%     \centering
%     \includegraphics[width=0.8\linewidth]{Figures/IMG/APP-PSF_IMG_all_defects.pdf}\vspace{-0.5cm}
%     %
%     \includegraphics[width=0.8\linewidth]{Figures/MICADO/MICADO_APP-PSF_all_defects.png}
%     \caption{This figure visualizes the APP-IMG (top row) and MICADO \acs{vAPP} (bottom row) performance with and without defects. The left column shows the idealized \acs{PSF}. A more realistic result that includes coating defects and write errors is presented in the middle column. Finally, the absolute difference in intensity is presented in the right column. Note that a nominal ($\pi$-valued) ideal half-wave optical retarder has been assumed even for the PSF with defects, but the spatially varying thickness variations were included.}
%     \label{fig:simulated-APP-with-defects}
% \end{figure}

% \begin{figure}[t!]
%     \centering
%     \includegraphics[width=0.75\linewidth]{Figures/APP_defects_contrast_realistic.pdf}
%     \caption{The simulated contrast curves for the METIS and MICADO \acs{vAPP}s, including various defect modes. The curves are an azimuthal average, computed from a large slice of the dark zone starting at the \acs{PSF} peak and measuring radially outwards to 25$\lambda/D$. The shaded areas illustrates the region between the \acs{IWA} and the \acs{OWA} of the respective \acs{vAPP}.}
%     \label{fig:APP-contrast-curves}
% \end{figure}

\begin{figure}[p]
\centering
    \captionsetup{width=\linewidth}
    \includegraphics[width=0.8\linewidth]{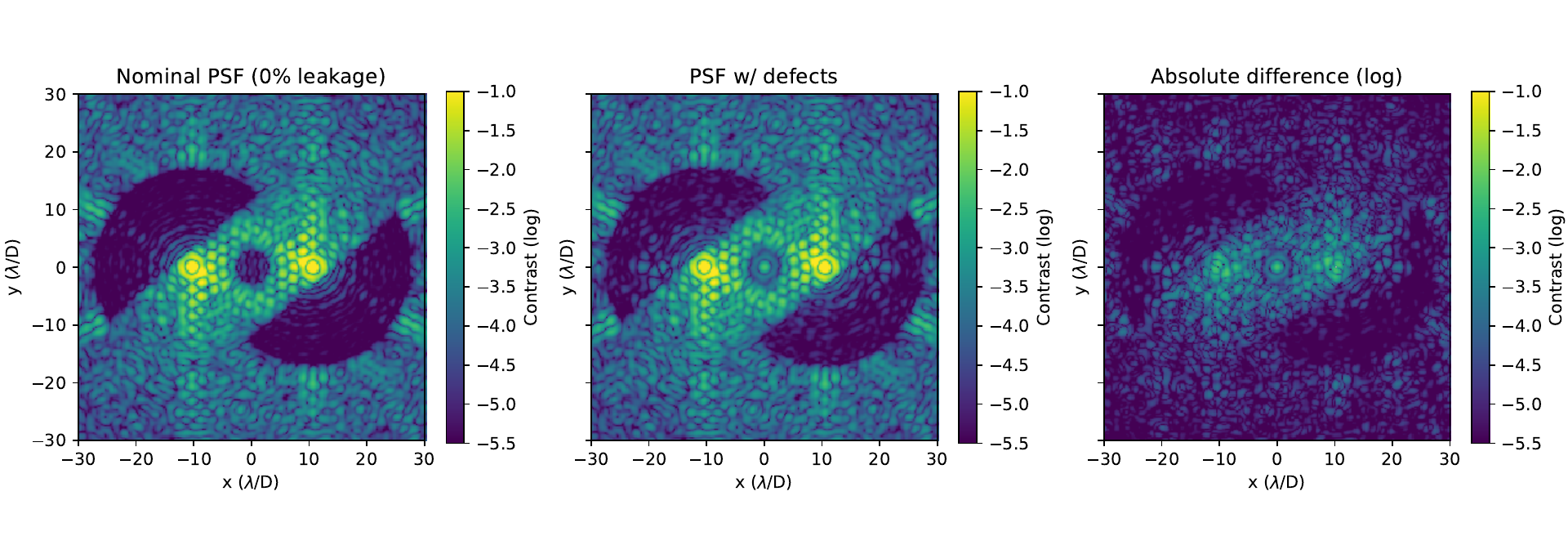}\vspace{-0.5cm}
    \includegraphics[width=0.8\linewidth]{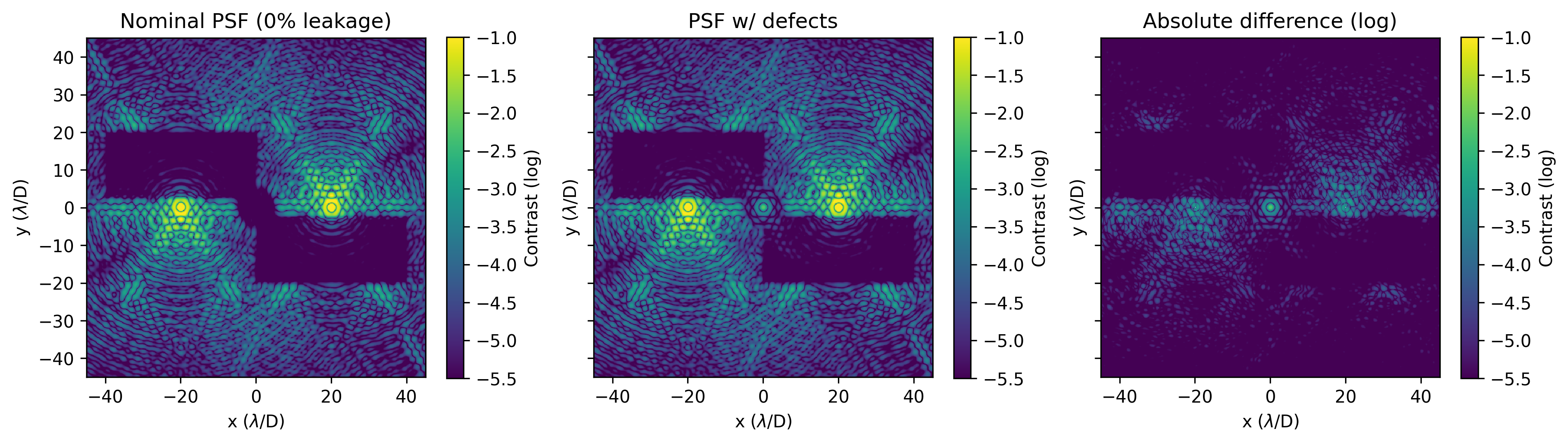}
    \captionof{figure}{This figure visualizes the APP-IMG (top row) and MICADO \acs{vAPP} (bottom row) performance with and without defects. The left column shows the idealized \acs{PSF}. A more realistic result that includes coating defects and write errors is presented in the middle column. Finally, the absolute difference in intensity is presented in the right column. Note that a nominal ideal half-wave optical retarder has been assumed even for the PSF with defects, but the spatially varying thickness variations were included.} \label{fig:simulated-APP-with-defects}
    \vspace{\baselineskip}
    \captionsetup{width=\linewidth}
    \includegraphics[width=0.7\linewidth]{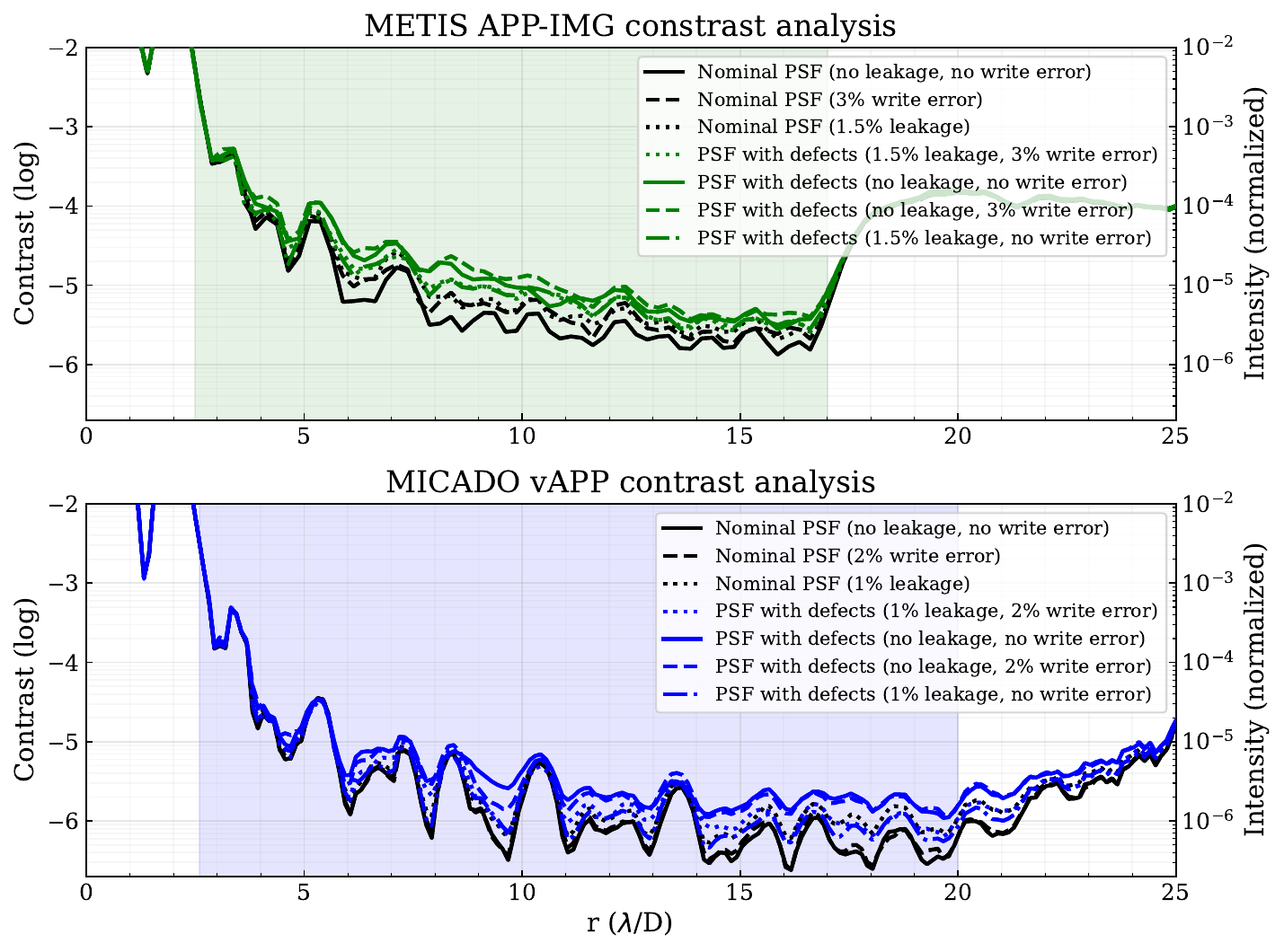}
    \captionof{figure}{The simulated contrast curves for the METIS and MICADO \acs{vAPP}s, including various defect modes. The curves are an azimuthal average, computed from a large slice of the dark zone starting at the \acs{PSF} peak and measuring radially outwards to 25$\lambda/D$. The shaded areas illustrates the region between the \acs{IWA} and the \acs{OWA} of the respective \acs{vAPP}.}
    \label{fig:APP-contrast-curves}
\end{figure}

An overview of the simulated defects and the assumptions we have made about their properties is provided in Table~\ref{tab:sim_defects_spec}. The magnitude, size and frequency of the defects is based on observations of the final METIS APP-IMG optic. Generally, we have tended towards conservative values where there was any doubt about their impact. Any wavefront errors originating from the coating and lamination processes are not included in this analysis. The results of the nominal (ideal) \acs{PSF} and the aberrated \acs{PSF} are shown in Fig.~\ref{fig:simulated-APP-with-defects}. Next, for a number of variations of defect complexity, the raw contrast curves are shown in Fig.~\ref{fig:APP-contrast-curves}. 

Through the obtained experience of the production of the METIS APP-IMG, various improvements were made for the production of the MICADO \acs{vAPP}. Because the optic is larger, contamination and smaller defects have smaller impact on the optic as they fill a substantially smaller part of the pattern. Furthermore, the coating deposition process could be better monitored, because the intended wavelength range was within the measurable range of the ellipsometer. Thus, a similar analysis as the one above carried out on the MICADO \acs{vAPP} could use slightly more optimistic values and logically comes with slightly better results. The resulting simulations are presented Figs.~\ref{fig:simulated-APP-with-defects} and \ref{fig:APP-contrast-curves} as well.

Unsurprisingly, worse defects lead to worse contrast performance. A deviation from an ideal half-wave retarder leads to light leakage and this explains a significant fraction of the decreased contrast levels in the dark zones around the \acs{PSF} core. Fast axis write errors do not have a major impact on the azimuthally averaged contrast, but they do create specific and predictable speckles at higher diffraction orders. For the METIS APP-IMG this is an important effect to note, because these speckles will appear in the dark zones. For the MICADO \acs{vAPP} \acs{PSF}, they fall outside the critical regions and are therefore less critical. If all discussed effects are included, the raw contrast levels can be decreased by 0.5-1.0 dex at radial distances larger than 6$\lambda/D$. Near the PSF core, there is no noticeable difference, although in reality some decrease in Strehl ratio seems likely. Nonetheless, being a pupil optic, both \acs{vAPP}s are quite robust to defects and an expectation of good performance seems warranted if the degradation of the wavefront can be kept under control.

%% file: 05_future_outlook.tex
\section{Conclusions and future outlook}\label{sec:outlook}

At present, of the three \acl{vAPP}s for the upcoming \acs{ELT}, only the METIS APP-IMG has been delivered and is ready for integration into the instrument. The other two are at different stages in the manufacturing process. The METIS APP-LMS, is currently being prepared for production at \acl{CLJ} For the MICADO \acs{vAPP}, the optic has received its phase pattern and is awaiting lamination to the large substrate, pending an improved adhesion procedure. Representative test samples are currently in production that will help to test, qualify and refine the lamination process with respect to the alignment of the two substrates and especially in terms of the transmitted wavefront error.

A final performance verification of the two larger coronagraphs will come once they are installed in METIS and MICADO, respectively. They are too large to easily test in a lab setting. However, the individual components that make up the optic were tested and provide confidence to their performance. Nonetheless, an end-to-end test of the APP-LMS optic on the VODCA test bench will provide the required data to demonstrate this before these instruments come online in a few years. 

The production of these optics have been the culmination of a tight collaboration between NOVA and \acl{CLJ} Some of the fundamental challenges have been tackled and slowly but surely the production of these liquid crystal based coronagraphs moves out of the research and development phase and matures into commercial manufacturing. For the optics described in this work, we have shown through detailed analyses and measurements that the optical characteristics are excellent and that the remaining coating imperfections will only have a small impact compared to the nominal design. We have shown that raw contrast levels around a few times 10\textsuperscript{-5} or better can be expected, in line with the requirements of these optics for both METIS and MICADO. 

In the past years, the focus in the high-contrast imaging community has shifted away from purely pupil-plane based coronagraphs towards those located in the purely in the focal plane, towards a combination of focal plane and pupil plane optics\cite{Por2020,Haffert2025}, or even without a coronagraph at all\cite{Bohn2020,Wagner2020,Sutlieff2026}. With focal plane coronagraphs much higher contrast levels can be achieved than with pupil-plane based optics alone, but tighter requirements on positioning, alignment and defects tolerances make these even more challenging to manufacture. Already promising steps have been made in recent months to produce vector vortex coronagraphs with good contrast and exceptionally low polarization leakage at visible wavelengths\cite{landman2026}. Such developments are critical to achieve the required 10\textsuperscript{-10} contrast levels that will enable the discovery and characterization of Earthlike planets around other stars, one of the main scientific objectives of the Habitable Worlds Observatory (HWO). Liquid crystal based coronagraphs are a promising pathway to achieving those objectives.

But until then, the \acl{vAPP}s that were the subject of this work will help in achieving the scientific objectives of the \acs{ELT}. In fact, in contrast to the strict pointing stability requirements of focal plane based coronagraphs, the robustness of pupil based coronagraphy will likely mean that some of the first direct detections and direct characterizations of exoplanets with the \acs{ELT} will be done with these optics. An exciting time to look forward to.